\documentclass[twocolumn]{aastex631}
\usepackage{enumitem}
\usepackage{courier}
\usepackage[T1]{fontenc}
\usepackage{epsfig}
\usepackage{epstopdf}
\usepackage{natbib}
\usepackage{amssymb}
\usepackage{amsbsy}
\usepackage{subfigure}
\usepackage[mathcal]{euscript}
\usepackage{float}
\usepackage{amsmath}
\usepackage{tabularx}
\usepackage{xspace}
\usepackage{enumitem}
\usepackage{stackengine}
\usepackage{comment}

\usepackage{ulem,xspace}

\newcommand{\msun}{M$_\sun$}
\newcommand{\kms}{km~s$^{-1}$}
\newcommand{\vsini}{\ensuremath{v \sin{i}}\xspace}
\newcommand{\kepler}{\textit{Kepler}}

\newcommand{\ali}{$A$(Li)}

\bibpunct{(}{)}{;}{a}{}{,}

\begin{document}
\title{Lithium in Kepler Red Giants: Defining Normal and Anomalous}

\author[0000-0002-4818-7885]{Jamie Tayar}
\affiliation{Department of Astronomy, University of Florida, Bryant Space Science Center, Stadium Road, Gainesville, FL 32611, USA }+
\affiliation{Institute for Astronomy, University of Hawai‘i at Mānoa, 2680 Woodlawn Drive, Honolulu, HI 96822, USA}

\author[0000-0001-5926-4471]{Joleen K.\ Carlberg}
\affiliation{Space Telescope Science Institute,
3700 San Martin Dr.,
Baltimore, MD 21218}

\author[0000-0002-9052-382X]{Claudia Aguilera-G\'omez}
\affiliation{Instituto de Astrof\'isica, Pontificia Universidad Cat\'olica de Chile, Av. Vicu\~na Mackenna 4860, 782-0436 Macul, Santiago, Chile}

\author[0000-0001-6180-8482]{Maryum Sayeed}
\affiliation{Department of Astronomy, Columbia University, 550 West 120th Street, New York, NY, USA}

\begin{abstract}
The {orders of magnitude variation in} lithium abundances of evolved stars have long been a puzzle. Diluted signals, ambiguous evolutionary states and unknown masses have made it challenging to both  map the expected lithium signals and explain the anomalously lithium-rich stars. We show here using a set of asteroseismically characterized evolved stars that the base lithium abundance in red giant stars is mass dependent, with higher mass stars having higher `normal' lithium abundances, while highly lithium enhanced stars may cluster around 0.8 or 1.8 \msun. We confirm previous studies that have shown that lithium enhancement and rapid rotation are often coincident, but find that the actual correlation between lithium abundance and the rotation rate, whether surface rotation, internal rotation, or radial differential rotation, is weak. Our data support previous assertions that most lithium rich giants are in the core-helium burning phase. We also note a tentative correlation between the highest lithium abundances  and unusual carbon to nitrogen ratios, which is suggestive of binary interactions, though we find no simple correlation between lithium richness and indicators of binarity.
\end{abstract}

\section{Introduction}

Recently, there has been renewed interest in the question of the lithium-rich giants. 
These puzzling stars show high lithium (Li) abundances, in excess of even the simplest models of Li destruction on the main sequence and subsequent dilution of the Li signal due to the deepening convective envelope on the red giant branch (RGB). Modern stellar evolution models and decades of observational Li measurements in red giants have shown that  simple models significantly underestimate Li destruction and dilution, making the Li-enriched red giants even more puzzling.

Large spectroscopic surveys including GALAH \citep{DeSilva2015}, LAMOST \citep{Cui2012}, and Gaia-ESO \citep{Gilmore2012} have allowed the identification of both lithium-rich - A(Li) $\gtrsim 1.8$ dex - and super-lithium-rich - A(Li) $ \gtrsim 3.2$ dex - giants \citep{DeepakReddy2019}. To define an enriched giant, the typical dredge-up dilution of a solar-type star is considered, while for the super Li-rich, the limit is based on the interstellar medium Li abundance value \citep{Knauth2003}. Based on that definition, \citet{DeepakReddy2019} have suggested that 0.6\% of stars are lithium enhanced and 0.04\% of stars are super lithium enhanced, similar to that of other studies \citep[][to name a few]{Brown1989,Kirby2012,Gao2019} confirming how rare these objects are. These limits to define Li enrichment are not strict and different works use different values. Additionally, authors have argued that the lithium abundances are mass, metallicity, and evolutionary state dependent \citep[e.g.,][]{AguileraGomez2016,Kumar2020}. This implies that the percentages of unusual giants can change when considering additional information or different limits \citep[e.g.,][]{Martell2021}.

Although many Li rich giants have been found, their origin remains a mystery. One of the hypotheses is internal lithium production through the eponymous Cameron-Fowler mechanism \citep{CameronFowler1971}, which requires the production of Be in the interior of the star. The difficulty for most first ascent red giants is that the Be can only be produced below the convection zone, requiring an unknown efficient mechanism to quickly transport the Be to the cooler convection zone, where it can increase the surface lithium and will not be destroyed, before it transforms into Li by electron capture. 
There are also some  suggested mechanisms of Li production during the He-flash or RGB tip \citep{Schwab2020, Mori2021}. Another set of explanations for Li-rich giants, especially those located before the luminosity function bump, is the pollution from an external source, such as substellar mass companions \citep{alexander1967, siessLivio1999}, or mass transfer from an asymptotic giant branch star which can produce Li by hot bottom burning \citep{SackmannBoothroyd1992}. It is also possible that the external source does not directly transfer Li to the giant but somehow triggers the Li production, such is the case of mergers of an RGB star with a helium-core white dwarf \citep{Zhang2020}, or potentially when a binary companion enhances rotation and rotational mixing, which can then mix Be up from the interior 
\citep{DenissenkovHerwig2004}.

Given the variety of theories, identifying the actual enrichment process for any unusual giant can be tricky, but important information is provided by their masses and exact evolutionary stages. Although it was first noted that most lithium-rich giants have surface gravities which would place them either in the core-helium burning red clump phase or close to the luminosity function bump \citep{Gratton2000}, new asteroseismic and spectroscopic measurements have allowed confirmation that a large fraction of the enriched giants are located in the red clump \citep[e.g.,][]{Yan2021, Martell2021, DeepakLambert2021a, DeepakLambert2021b, Singh2019, Minghao2021}. The mass dependence of lithium abundances, on the other hand, is still very much an open question. It appears that there is a mass dependent Li depletion on the main sequence that changes the abundances stars have when they enter the RGB phase \citep[see e.g.][]{SestitoRandich2005}. Then, the first dredge-up dilution and possible additional transport processes (such as thermohaline mixing) in the upper RGB are also mass dependent \citep{Magrini2021a}. 

Moreover, considering that stars of different masses may have different initial Li abundances at formation, mass is definitely something to consider when analyzing the Li pattern of stars. Even if the probability of enrichment is independent of stellar mass, as suggested by \citet{Deepak2020}, given the underlying distribution of the population, we expect to find a different amount of enriched giants at different masses and a different {criterion} to define unusual giants for each mass. In addition, it has been argued that the lithium richness in first ascent red giant branch stars should not exceed A(Li)$=2.6 \pm 0.24$ dex, a limit observed in a sample of well-characterized RGB and red clump giants \citep{Yan2021}.

Another ingredient to consider is the possible correlation between other signatures and Li abundance. In particular, rapidly rotating stars are more likely to have detectable lithium \citep[e.g.][]{Drake2002}. Related to this particular observational signal, it has been argued that the enhancement of Li on the red clump and its correlation with rapid rotation implies a mechanism whereby stars are enhanced by rotational mixing driven by tidal interactions near the tip of the red giant branch \citep{Casey2019}. The rapid rotation of Li enriched giants has also been associated with a planet engulfment event \citep{Carlberg2012}, that could produce both observational signatures at the same time.

{Significant work has been done to look at the lithium abundances of red giants in clusters, and to look for correlations with mass, rotation, metallicity, binarity, and evolutionary state \citep[e.g.][]{Carlberg2016, DelgadoMena2016, AnthonyTwarog2021, Sun2022}. However, there are only a limited number of bright giants in nearby clusters, and given the number of potentially relevant parameters, such work has been challenging. The studies of Li abundance in the much larger sample of field RGB stars} are often complicated by the lack of directly measured masses and evolutionary stages for the stars of interest, where these parameters had to be inferred from HR diagram position. This is challenging in this regime where the evolutionary tracks bunch together. However, asteroseismology, the study of stellar oscillations, allows the measurement of stellar mass from the calibrated combination of the frequency of maximum power of the oscillations and the large frequency separation \citep{KjeldsenBedding1995}. In addition, the energy generation in the core changes the structure of the interior and thus the sound speed profile, making it possible to directly estimate the evolutionary state from the mixed mode pattern \citep{Bedding2011, Mosser2014, Elsworth2019}. Asteroseismology thus allows the measurement of both the evolutionary state and the stellar mass, and can therefore help illuminate the Li pattern. 


In this paper, we use these direct asteroseismic measurements to test the recent inferences of the Li pattern's dependence on mass and evolutionary state in a carefully chosen set of stars with known stellar parameters. We focus on stars that are more metal rich, which allows us to better understand the mass dependence at a specific metallicity where most of the Li-rich giants seem to be found, where the effect of extra-mixing in the RGB is weaker \citep{Shetrone2019, Lagarde2019, AguileraGomez2022}, and where the close binary fraction is lower \citep{Badenes2018, Moe2019}. Moreover, with asteroseismology, we can now study possible correlations of Li with core rotation and the rotation profile, better constraining possible enrichment mechanisms.

\section{Sample Selection} \label{Sec:sample}


\begin{figure}[htb]
\begin{center}
\subfigure{\includegraphics[width=8.5cm,clip=true, trim=0.5in 0.1in 0in 0.4in]{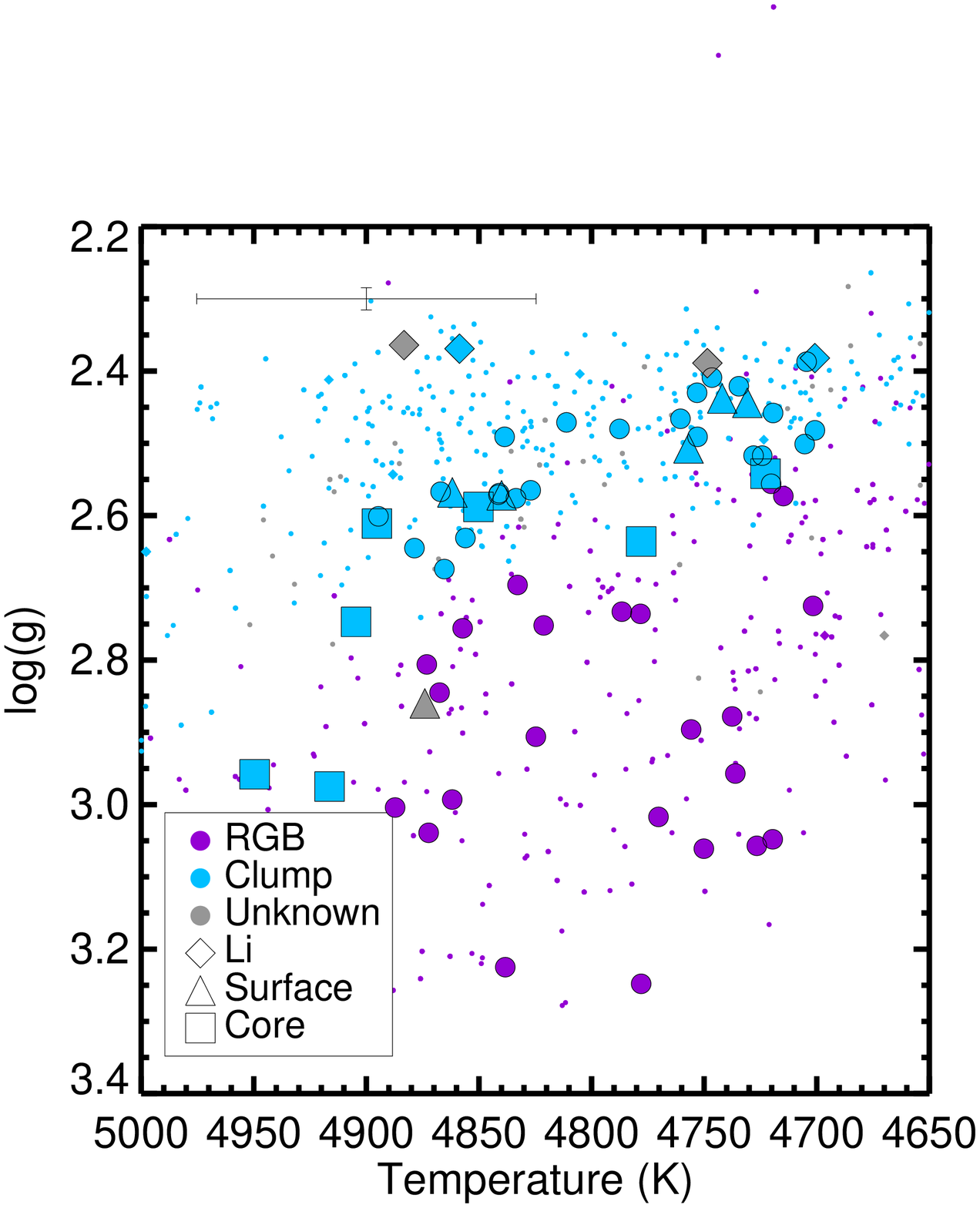}}
\subfigure{\includegraphics[width=8.5cm,clip=true, trim=0.5in 0.2in 0in 0.4in]{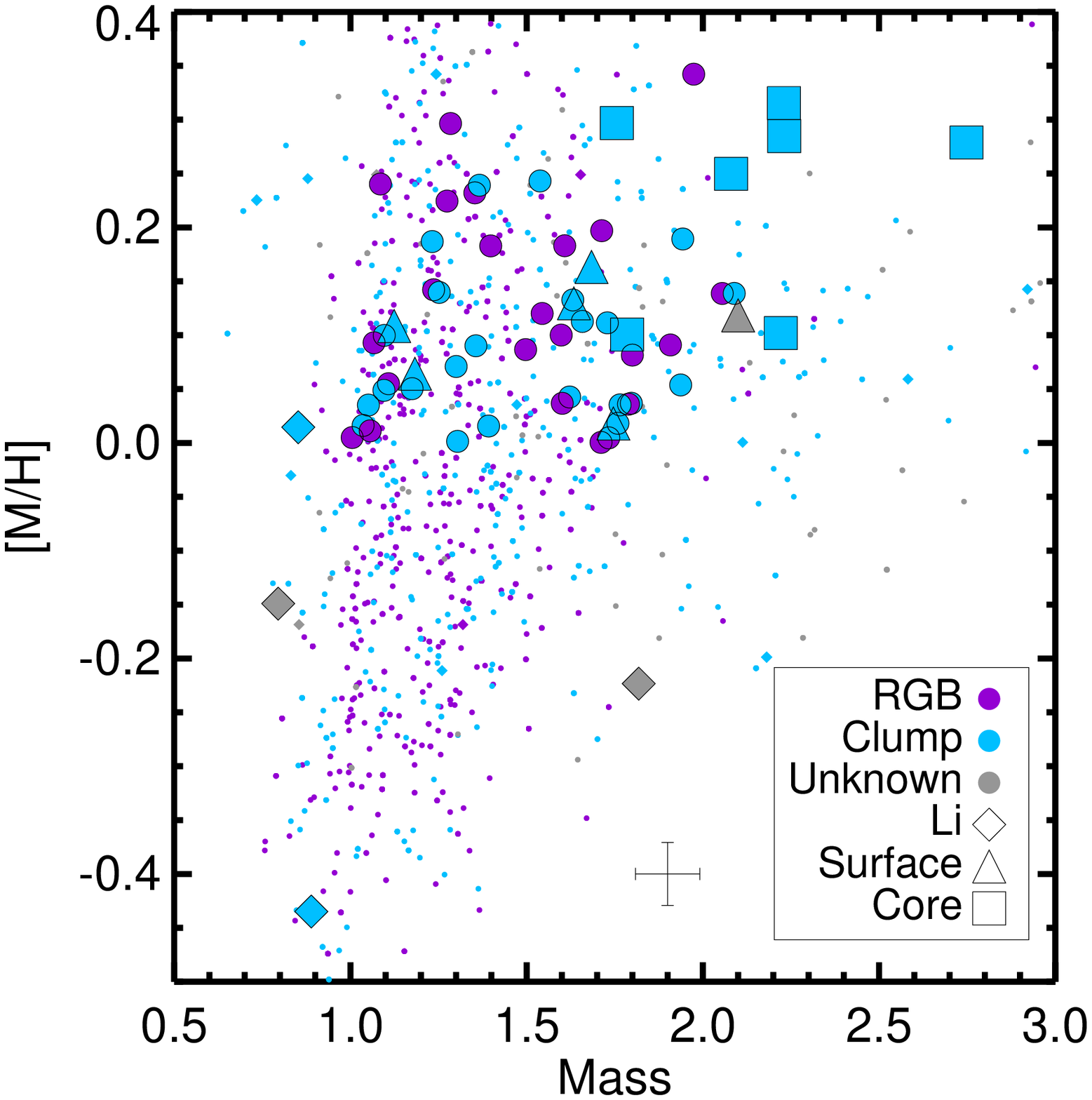}}
\caption{Our sample, with the color indicating the seismic evolutionary state, and the shape of the point indicating the reason the star was selected (circle: mass bins; diamond: identified as lithium-rich by \citet{Casey2019}; triangle: measurable surface rotation; square: part of the core rotation sample of \citet{Tayar2019b}). For comparison we have shown the APOKASC sample's overlap with LAMOST data \citep{Gao2021} as small points in the background.}

\label{Fig:kiel}
\end{center}
\end{figure}

In this study, we select stars of known mass and evolutionary state for analysis. We use as a basis for selection the APOGEE-\kepler\ catalog \citep{Pinsonneault2018}. These are stars with asteroseismic parameters from the analysis of \kepler\ data \citep{Borucki2010} by five different asteroseismic pipelines, which include theoretical corrections to the scaling relations \citep{White2011,Sharma2016,Pinsonneault2018}, as well as empirical corrections to match the mass scale of the clusters in the \kepler\ field. The estimation of evolutionary states from an ensemble of methods is described in detail in \citet{Elsworth2019}. 

These stars also have detailed spectroscopic characterization from the APOGEE survey \citep{Majewski2017}, which is a Sloan Digital Sky Survey IV \citep{Blanton2017} program using the 2.5-meter telescope \citep{Gunn2006} with the APOGEE spectrograph \citep{Wilson2019} to collect spectra of stars at moderate resolution (R $\sim$ 22,000) in the H-band. The data are reduced using the ASPCAP pipeline \citep{Nidever2015,GarciaPerez2016} and calibrated to clusters \citep{Meszaros2013} as well as asteroseismic data \citep{Pinsonneault2018}. For our study, we use the Data Release 14 data \citep{DR14} whose analysis is discussed in detail in \citet{Holtzman2018}. We make this choice for consistency with the APOKASC-2 analysis \citep{Pinsonneault2018} and our target selection, although more recent data releases are now available \citep{DR16, DR17}.

{For our analysis, we collected additional, higher resolution spectra for a limited subsample of the APOGEE-\kepler\ stars. The strength of the lithium line is known to be correlated with temperature, gravity, metallicity, as well as actual lithium abundance, which is correlated with mass, metallicity, evolutionary state. Since we knew our sample would be too small to simultaneously account for all of these variables, we made choices to reduce the number of axes of variation included in our study. Specifically, we restricted ourselves to a limited metallicity range and a limited range in temperature. Our initial sample had very strict cuts on both metallicity and effective temperature (\ref{sec:sample_mass}), but to collect a sufficient sample for cross-validation (\ref{ssec:crossvalidation}) and to compare to internal rotation rates (\ref{ssec:corerotselect}) we were forced to relax these limits slightly. The exact cuts used to select each sample are documented in the appropriate sections. }

\subsection{Mass Bins}
\label{sec:sample_mass}
The core of this analysis is the selection of stars in mass bins so we restrict the sample to narrow bins in other stellar parameters. 
The strengths of the strong Li lines are very sensitive to temperature and are strongest at cooler temperatures for a fixed Li abundance; therefore, we restrict our sample to the region around 4800 K {($\pm$100 K,} see Figure \ref{Fig:kiel}). 
In addition, several authors have suggested a metallicity dependence to the lithium distribution \citep[e.g.][]{Martell2021}. To remove this axis of variability, we also choose stars with metallicities between 0.0 and +0.4 dex. We then divide the stars using their asteroseismically measured evolutionary states \citep{Elsworth2019} into core-helium-burning clump stars and shell-hydrogen-burning first ascent red giants. 

Within each sample, we divide the stars between 0.9 and 2.1 solar masses into six mass bins, and identify the brightest stars in each bin. We were able to observe 23 giants and 23 clump stars, which gave us between three {and} five stars in each bin. As shown in Figure \ref{Fig:kiel}, requiring stars at the same temperature but different evolutionary states means that there is a slight offset in surface gravity between our stars in the core-helium-burning phase and those in the shell-hydrogen-burning phase; we do not expect this offset to substantially affect our analysis. We note that at the chosen temperature, the vast majority of the red giant branch stars in our sample should be below the red giant branch bump, 
which happens around a surface gravity of 2.65 dex for stars at this metallicity, but all should have completed their first dredge up. 



\subsection{External Cross-validation}
\label{ssec:crossvalidation}
Recent work by e.g. \citet{Casey2019} has attempted to identify Li-rich giants and even measure their Li richness from low-resolution (R $\sim$ 10,000) spectra from LAMOST \citep[Large Sky Area Multi-Object Fiber Spectroscopic Telescope;][]{LAMOST1}. In order to validate both our abundances and the work being done with lower resolution measurements, we include four stars with asteroseismic measurements that meet our original temperature criteria, but were too faint and/or metal-poor and would not have otherwise been selected. We note that in the course of our analysis, \citet{Gao2021} published Li abundances for several additional stars in our sample and we add those results to our comparison to LAMOST. For illustrative purposes, and to put our smaller sample in context, we also occasionally compare our results to various subsets of the full overlap sample between \citet{Gao2021} and APOKASC-2 \citep{Pinsonneault2018} throughout this paper. Therefore, we have also shown the \citet{Gao2021} sample as small points in the background of several of our plots for reference, including Figure \ref{Fig:kiel}.

 
\subsection{Rapid Rotation}
There are well established connections between rapid rotation and Li enhancement \citep[e.g.][]{Carlberg2012}. In addition, in conjunction with mass measurements, rapid rotation has been used as an indicator of potential binary interactions in red giants \citep{Tayar2015, Ceillier2017, Daher2022}. To test these connections and their relationship with stellar mass, we added to our sample one giant with a measured spectroscopic rotation velocity and five giants with spot modulation periods \citep{Ceillier2017}
within or just outside our selection criteria. This will allow us to test the mass, evolutionary state, and rotation dependence of the lithium abundance or upper limits at fixed metallicity and sensitivity. We note that several other stars in our sample either had measurements of rotation or had useful upper limits on their rotation velocity inferred from our spectroscopic analysis and we add those to our analysis as well.

\subsection{Core Rotation}
\label{ssec:corerotselect}

While previous work has looked for correlations between surface rotation and Li enhancement, in many cases it is the rotation rate of the interior of the star that should provide insight into the connection between rotation and mixing. With asteroseismology, it has become possible to measure the core rotation rates of evolved stars \citep[e.g.][]{Beck2011, Mosser2012b}. We add to our sample a set of seven more massive stars with measured core rotation rates from \citet{Tayar2019b} that are close to 4800 K {($\pm$200 K)}. We also note that after our selection, we discovered that an additional five giants in our sample have measured core rotation rates from \citet{Gehan2018}, and so we include that information in our analysis. Our sample selection is documented in Table \ref{Tab:table1}. 



\section{Observations and Data Reduction}
Our higher resolution optical spectra were taken with the High Dispersion Spectrograph \citep[HDS;][]{Noguchi2002} on the Subaru telescope \citep{Iye2004} {on July 9th and 10th, 2019. The stars ranged in Kepler magnitude from 9 to 12.} A non-standard setup using a cross scan 
rotation angle of $4.533^\circ$ and a grating 
angle of $0.2408^\circ$ yielded spectral coverage of $\sim$5700--7050~\AA\ in the red arm. Only data taken from the red side were reduced and used in this analysis. Data were taken with the 2.0''x30'' slit and utilized the 0.2''x3 image slicer. The image slicer allows one to potentially reach higher resolving powers of narrower slits with less of a penalty to light lost in typical seeing conditions. However, because the intrinsic broadening of red giants nullifies the benefits of such high spectral resolution, we opted to extract the image slices with a single long aperture in IRAF \citep{Tody1986IRAF,Tody1993IRAF}. From measuring the width of ThAr lines, we find that our spectra have a typical resolving power of $R\sim85,000$ (3.5~\kms).


The HDS spectra were reduced with IRAF, following the reduction guidance in the HDS IRAF Reduction Manual \footnote{\url{https://www.subarutelescope.org/Observing/Instruments/HDS/specana2014.10e.pdf}}. Specialized routines for the overscan correction and non-linearity correction of the data, available from the HDS website\footnote{\url{https://www.subarutelescope.org/Observing/Instruments/HDS/}}, were used in the reduction. Additionally, standard bias subtraction, flat fielding, cosmic ray removal, and scattered light removal were performed. 
The wavelength solutions were measured from ThAr comparison lamp spectra taken throughout the night, and the solution for each stellar spectrum was interpolated from the comparison lamp spectra taken at the nearest time.  The continuum of each echelle order was fit 
and then divided out by the blaze function. The echelle orders were then intercombined using \emph{scombine} to create the final one-dimensional spectra.  These spectra were cross-correlated with the \cite{Hinkle2005} atlas Arcturus spectrum to measure the observed radial velocity, which was used to shift the final spectra to the stellar rest frame.

\section{Stellar Characterization} \label{Sec:Characterization}
\subsection{Lithium Measurements}

\begin{figure}
    \centering
    \includegraphics[width=0.4\textwidth,angle=90]{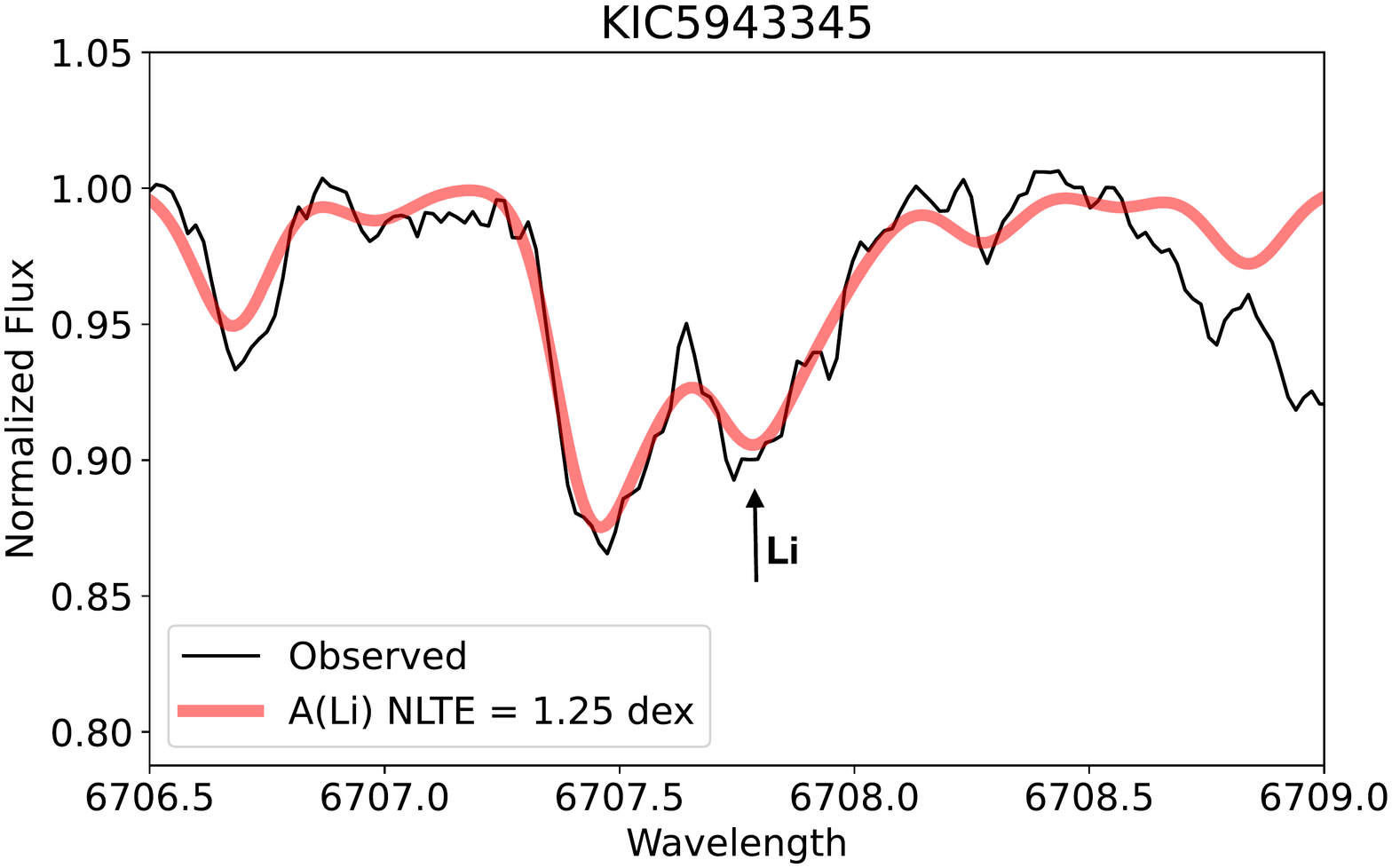}
    \includegraphics[width=0.48\textwidth]{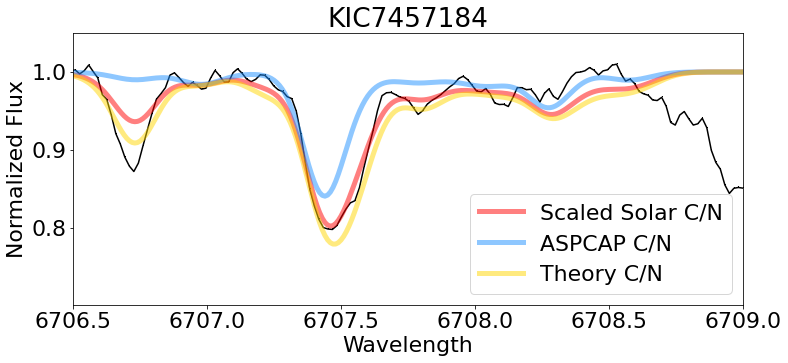}
    \caption{{\textbf{Top:} Example best fit spectrum (red) to the observed spectrum (black) of KIC5943345. The Li line is centered in this plot and is the second strongest feature for this star in this wavelength range. \textbf{Bottom:} Our analysis uses scaled-solar (red) CN abundances rather than theoretical (yellow) or directly measured ones (blue) as empirically they seem to provide better fits.}}
    \label{fig:ex_li_fit}
    
\end{figure}
The abundances of Li were measured via synthetic spectrum fitting of the resonance lines in a small bandpass between 6706 and 6709~\AA, using the 2019 Version of MOOG\footnote{Downloaded from \url{https://www.as.utexas.edu/~chris/moog.html}} \citep{Sneden1973}. In cases of high Li abundance, we also checked the measurement for consistency with the subordinate Li lines at 6104~\AA. The line list to generate the spectra draws the atomic information from \cite{Ghezzi2009} and replaces the carbon and nitrogen (CN) data in that work with the new linelists from \cite{Sneden2014}. 
Atmosphere models for each star were interpolated from the grid of MARCS spherical atmosphere models \citep{Plez2008}, using the stellar parameters derived from APOKASC \citep{Pinsonneault2018}, specifically, the corrected temperature, 
the asteroseismic $\log g$, and the stellar metallicity. A single microturbulence value of 1.5~\kms\ was used for all stars. To account for the effects of first dredge-up, the carbon and nitrogen abundances were forced to have a ratio of 1.5 while preserving the original scaled-solar sum total number abundance. 

A number of atomic features as well as CN molecular features have a large impact on the spectrum in the bandpass of interest, and we tested three different ways of modeling the region.  We first generated synthetic spectra by assuming all stars have scaled solar abundances of all elements (other than the C/N adjustment). We then generated a second batch of synthetic spectra where we adopt the APOGEE-measured abundances of C, N, Si, V, Ca, and Fe with the expectation that the atomic and molecular features would be fine-tuned to the star's individual chemical peculiarities. Finally, we used theoretically predicted C and N abundances from models appropriate to our stars \citep{Tayar2017}. We found cases where using these ``fine-tuned'' abundances (both APOGEE-measured and theoretical) significantly overestimated or underestimated features nearby the Li lines. Conversely, the scaled solar abundances, while not always the ``best'' fit to the non-Li features, were more consistently well-fit, and the scaled solar synthetic spectra were ultimately adopted for all of our fits (See Figure \ref{fig:ex_li_fit}). 
\begin{figure*}[bt!]
\begin{minipage}{1.0\textwidth}
\begin{center}
\subfigure{\includegraphics[width=8.5cm,clip=true, trim=0.5in 0in 0in 0in]{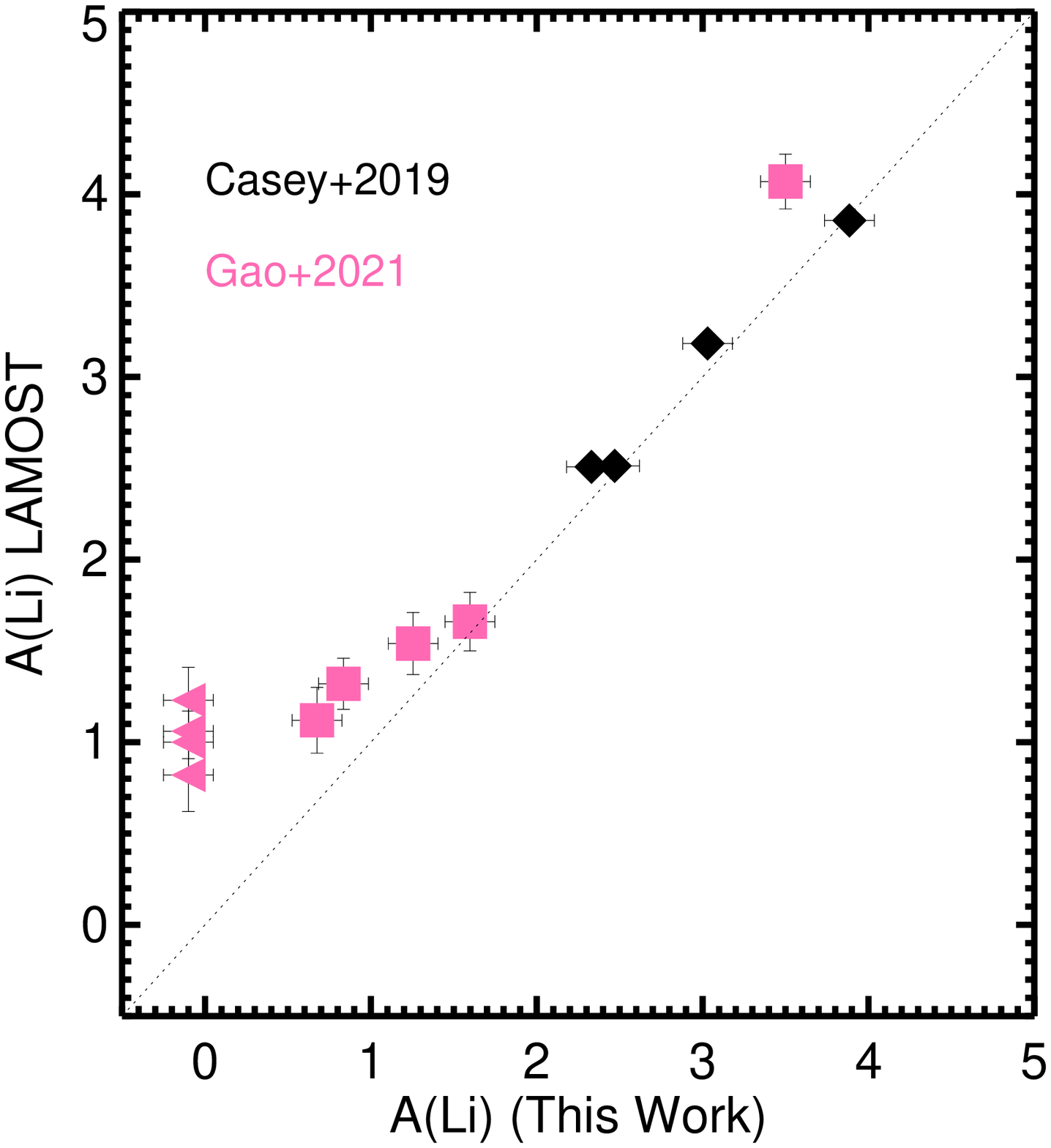}}
\subfigure{\includegraphics[width=8.5cm,clip=true, trim=0.5in 0in 0in 0in]{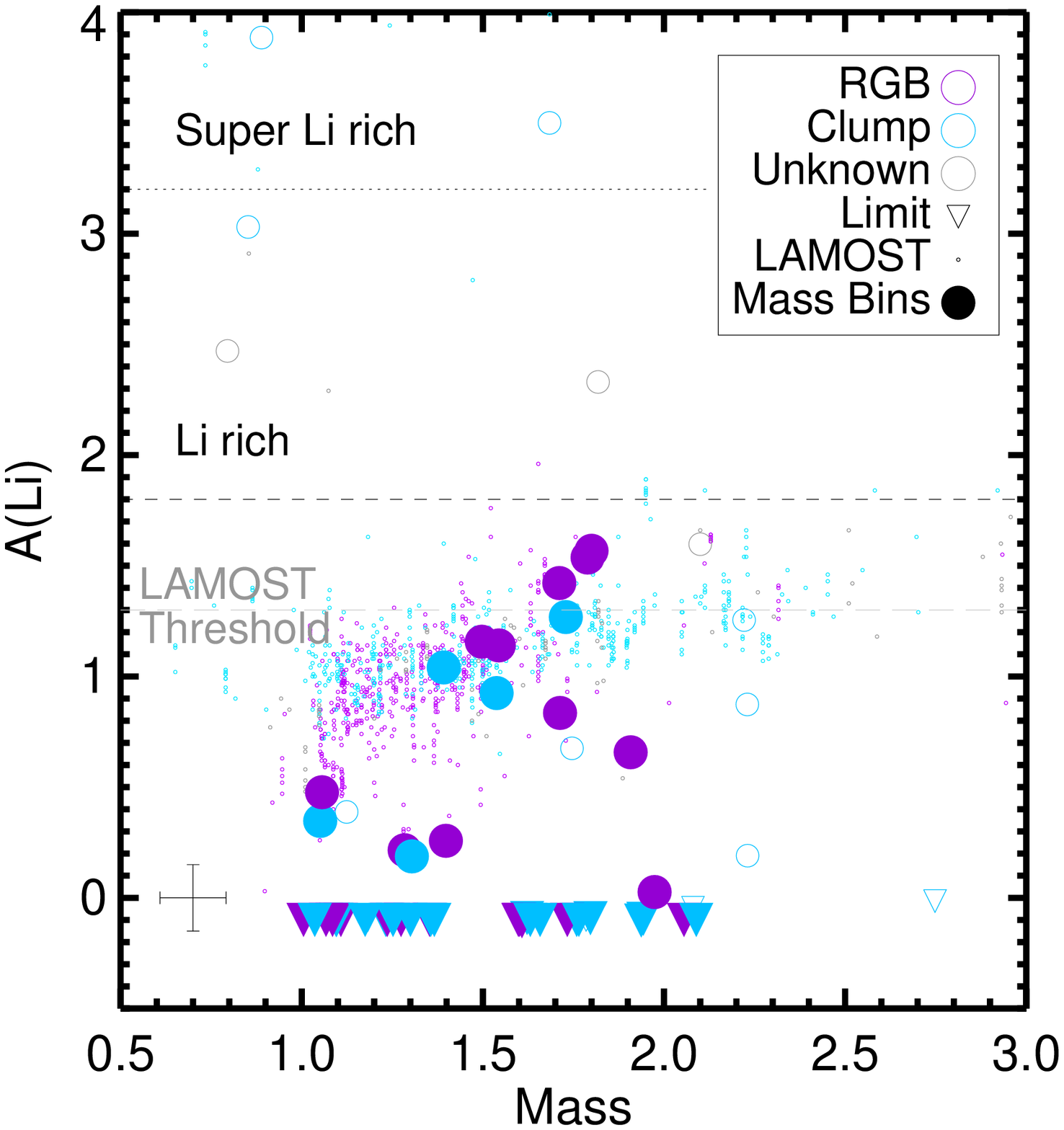}}
\caption{\textbf{Left:} Comparison of the lithium abundances measured in this analysis and the results published using  lower-resolution LAMOST spectra from  \citet{Casey2019} and \citet{Gao2021}. The correlation is quite strong, especially at abundances above A(Li) $\sim$ 1.3 dex, suggesting that LAMOST spectra are sufficient to identify truly lithium-rich giants. We note that our abundances include NLTE corrections whereas the LAMOST values both assume LTE, which could explain some small offsets. \textbf{Right:} Measured lithium abundances and upper limits for the RGB (purple) and clump samples (blue) as well as stars with ambiguous evolutionary states (grey). Stars included as part of the mass sample are shown as large filled symbols, whereas stars included for other reasons are shown as smaller open symbols. The thresholds for lithium richness and super-lithium richness from \citet{DeepakReddy2019} are shown for reference. For comparison, published values from LAMOST for stars with metallicities between $0.0-0.4$ dex are shown as tiny open circles; these do not have the same temperature restrictions as our sample. We also have concerns about the LAMOST measurement accuracy for stars below A(Li)$\sim$1.3, marked as `LAMOST Threshold', coming from the left panel.}

\label{Fig:casey}
\end{center}
\end{minipage}
\end{figure*}

The adopted broadening is another important factor for the spectral synthesis. The majority of the stars are slow rotators, and the total broadening can be well approximated by a simple Gaussian. To account for slight star-to-star variations in broadening, the neighboring isolated \ion{Fe}{1} line at 6750.15~\AA\ was fit with a Gaussian, and the associated full-width at half maximum (FWHM) was adopted for the Li fitting. For the three stars with a previous measurement of $\vsini > 6$~\kms, we compute separate broadening parameters, using the median FWHM as the instrumental plus macroturbulent velocity component, and the known $\vsini$ (adopting a limb darkening of 0.6) to model the rotation. For 18 stars ($\sim$ 30\% of the sample), the adopted FHWM was later reduced to improve the fits.

\begin{figure*}[t!]
\begin{minipage}{1.0\textwidth}
\begin{center}
\subfigure{\includegraphics[width=8.5cm,clip=true]{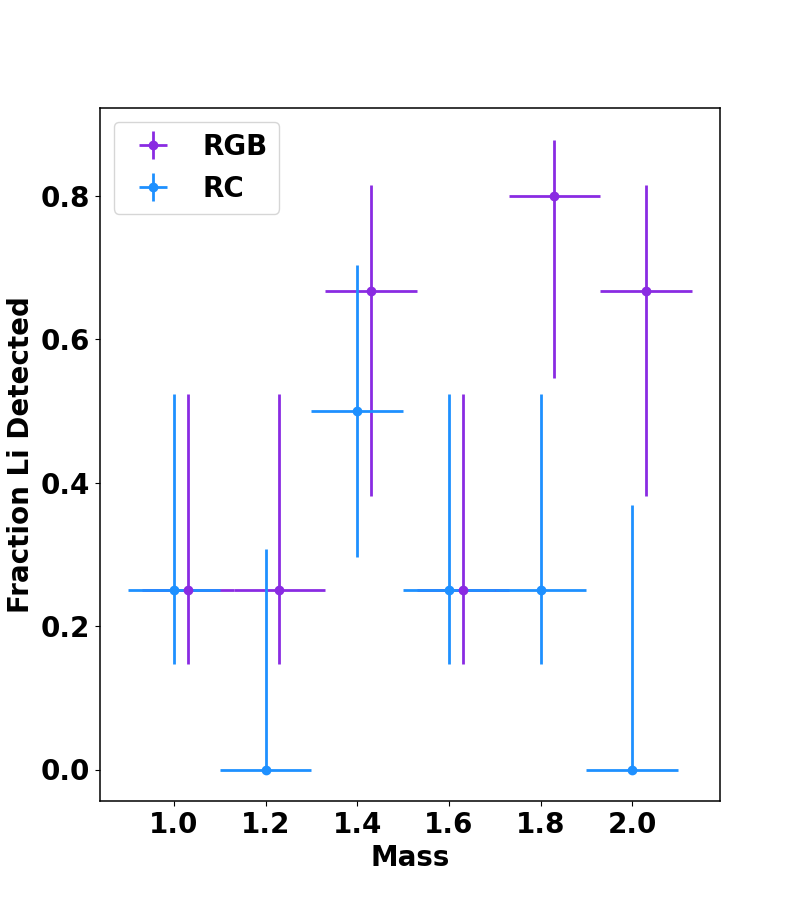}}
\subfigure{\includegraphics[width=8.5cm,clip=true, trim=0.5in 0in 0in 0in]{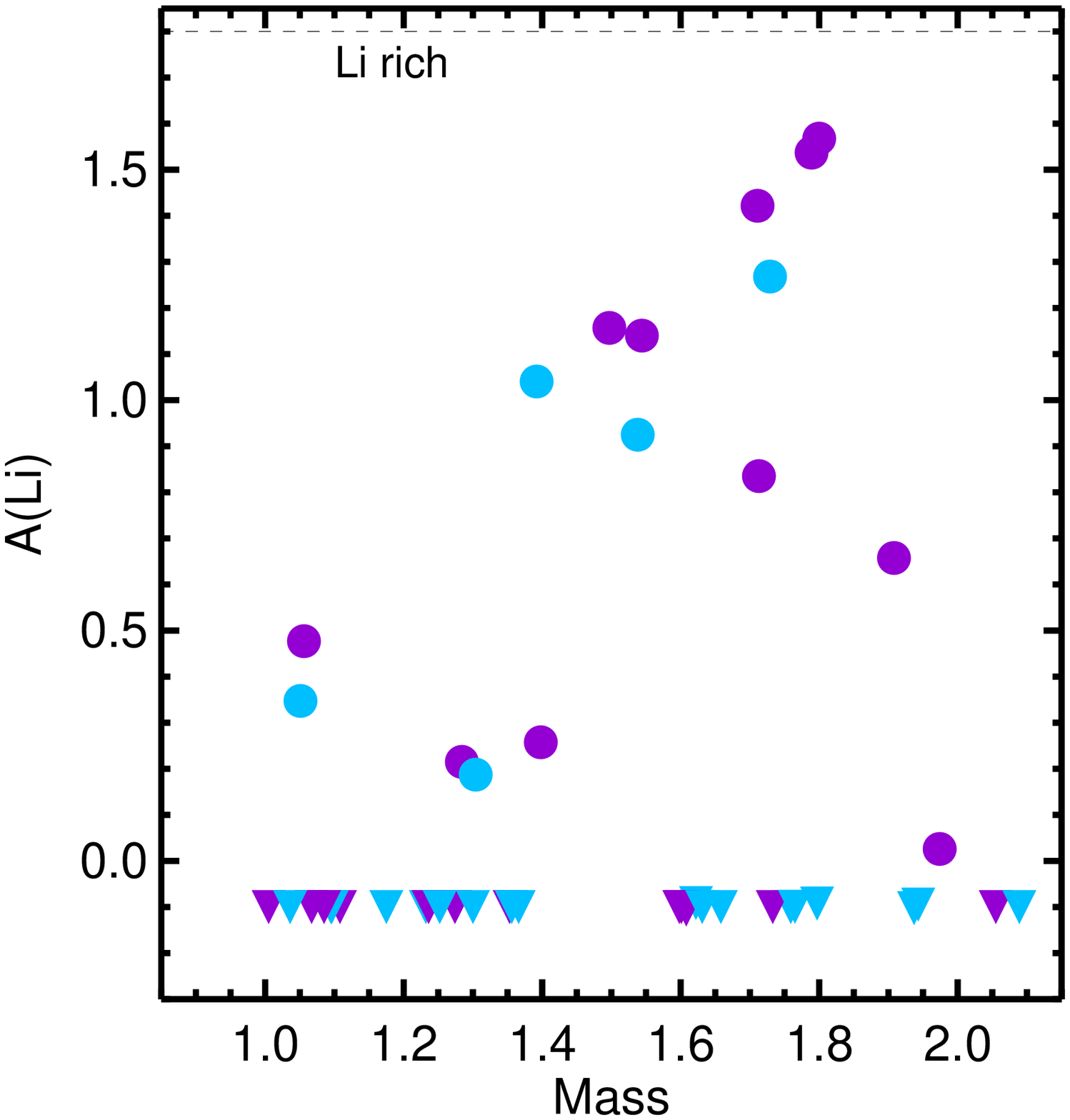}}

\caption{\textbf{Left:} Lithium detection fractions in bins of mass and evolutionary state (RGB in purple and clump in blue), {for stars selected along these parameters (Section \ref{sec:sample_mass}). The RGB points are offset by 0.03 $M_\odot$ from the bin center for clarity. Error bars are binomial confidence intervals equivalent to 1$\sigma$ probabilities.}
\textbf{Right:} Similar to the right panel of Figure \ref{Fig:casey} except that here we show only the measurements and upper limits for the stars chosen as part of the mass and evolutionary state selected sample shown in the left panel. }

\label{Fig:massbins}
\end{center}
\end{minipage}
\end{figure*}
The fitting procedure was semi-automated. All stars were initially run through an automated procedure where a custom-Python code ran a series of MOOG-generated synthetic spectra with a range of input Li-abundances. The MOOG parameter files for spectral synthesis work with $\Delta$ abundances from the star's atmosphere file, and the initial set of synthesized $\Delta$\ali\ are -3, -2, and -1.  The $\chi^2$ is computed for the model fits to the data in the narrow wavelength region centered on the Li-lines for the three models, and the $\Delta$\ali\ with the minimum $\chi^2$ is used as the next best guess in the following iteration.  Each iteration uses the best guess plus or minus a step size. When the minimum $\chi^2$ is associated with either the minimum or maximum $\Delta$\ali, that $\Delta$\ali\ becomes the new best guess, and the step size remains the same.  When the middle $\Delta$\ali\ of the current iteration has the minimum $\chi^2$,  a quadratic interpolation of the three pairs defines the next best guess, and the step size decreases in the following progression: 1.0 (the original step size), 0.8, 0.5, 0.2, 0.1, 0.05, and 0.01 dex. This process continues until either  a solution is reached (the code reaches the smallest step size) or an upper limit is detected (when the difference between the synthetic models tried in the iteration becomes less than a 1~m\AA). 
Figure \ref{fig:ex_li_fit} shows an example of one of the stars that was fit the automated procedure.

All of the fits from the automated run were visually inspected, and nine stars were identified at this stage as having poor quality spectra for Li measurements. An additional 16 stars had broadening adjusted to improve the quality of the fits and were rerun through the automated fitting. Four stars (all Li-rich) were fit fully by hand, trading off small variations in the line FWHM parameter and best-fit abundance. At these large line strengths, small errors in the broadening will lead to proportionally larger errors in the abundances.
{The final results are presented in Table \ref{Tab:table1}.}
We define our detection limit to be at a feature size of $\sim 5$~m\AA, which corresponds to an upper limit of \ali~$\sim +0.1$ dex. Typical uncertainties in \ali\ measurement come from the combination of fitting uncertainties ($\sim$ 0.05--0.1 dex) and propagated errors from stellar parameter uncertainties, which is dominated by the temperature uncertainty. For a temperature uncertainty of 100~K, the associated error in \ali\ is 0.13 dex. We adopt a typical uncertainty of 0.15~dex or our sample. Finally, non-local thermodynamic equilibrium (NLTE) corrections were applied. The NLTE corrections were interpolated from the \cite{Lind2009} grid of corrections.

\subsection{\vsini}
We measured \vsini\ from the spectra using a procedure similar to that used by \citet{Carlberg2014}, where the stars had been cross-correlated with radial velocity standard stars  broadened with a grid of rotational velocities. Because we did not observe any standard stars with HDS, we modified the procedure to instead measure \vsini\ using auto-correlation functions. We identified 16 wavelength bins, each 50~\AA\ wide, that are free of strong telluric absorption. We compute the autocorrelation function (ACF) for each star and each bin and record the FWHM of a Gaussian fit to the cross-correlation peak. The star with the smallest average FWHM (KIC~6103934) is selected as a representative slow rotator.  Its spectrum is broadened with a range of rotational velocity kernels from 1--6~\kms\ in 1~\kms\ steps, and from 6--26~\kms\ in 2~\kms\ steps.  The ACF fitting procedure is repeated at each rotational velocity, creating a mapping between input rotational velocity and fit FWHM for each wavelength bin. These FWHM-\vsini\ relationships are interpolated at the measured FWHM for each of the other science targets to estimate the \vsini, yielding 16 \vsini\ measurements per star. In Table \ref{Tab:table1} we report the measured \vsini\ (from the mean) and uncertainty (from the standard deviation). 

Inherent in this analysis is the assumption that all other broadening contributions (instrumental and macroturbulent) are constant across the sample. Additionally, the macroturbulent velocity of these class-III red giants is typically $\sim5$~\kms\ \citep{Gray2005}, larger than 
the \vsini\ of many of the slowest rotators.
This is why artificially broadening the spectra with input \vsini $\lesssim$ 3\--4~\kms\ has little effect on the measured FWHM. In fact, this method resulted in a measured \vsini\ of 3.9~\kms\ for the star selected as the likely smallest broadening. Therefore, we expect our method is unable to recover any \vsini\ below 4\--5~\kms. Such limits are consistent with the lack of detected rotational broadening in APOGEE for most of our stars \citep{Tayar2015, Dixon2020, Daher2022, Patton2023}.


\section{Analysis} \label{Sec:Analysis}
\subsection{Validation of Results at Low Resolution}

In Figure \ref{Fig:casey} we compare our measured Li abundances from our  analysis of  high-resolution Subaru spectra to those estimated from LAMOST low-resolution \citep{Casey2019} or medium-resolution \citep{Gao2021} spectra. In general, we find that LAMOST data was entirely sufficient for identifying Li-rich giants. We therefore suggest that with appropriate caution or calibration, they can indeed be used to identify large numbers of Li-rich giants across the galaxy. We do note however that as the Li abundance falls below about A(Li) $\sim$ 1.3 dex, the estimates based on lower-resolution data start to deviate from what we estimate from our higher resolution spectra, suggesting that the detection threshold in LAMOST is slightly underestimated for these cool, high-metallicity giants.  This limit is not surprising, since it is approximately where the Li feature becomes weaker than the neighboring Fe-dominated feature near 6707.5~\AA, as seen in Figure \ref{fig:ex_li_fit}. 
{In the right panel of  Figure \ref{Fig:casey}, we plot Li as a function of mass for both our sample and the LAMOST sample. While mass trends are discussed in detail in next section, we note here that the vast majority of LAMOST detections fall below this 1.3~dex threshold and are suspect. Nevertheless, the LAMOST measurements in general show similar trends to our own, with higher lithium abundances for more massive stars (i.e., stars more massive than $\sim$1.8M$_\sun$ tend to be above the LAMOST threshold) and a tentative preponderance of Li-rich red giants at low masses ($\sim$0.8M$_\sun$).}



\subsection{Correlations with Mass}

Li abundances on the red giant branch are sensitive to a wide range of complicated mixing processes that happen in earlier phases of evolution. One of the core motivations of our analysis was to establish a baseline for normal Li abundances as a function of stellar mass and evolutionary state, so that anomalous Li abundances can be more sensitively identified. {In the left panel of Figure \ref{Fig:massbins}, we show the fraction of stars with lithium above our detection threshold value as a function of stellar mass and evolutionary state using only the stars selected without reference to their rotation or Li enhanced status. Our detection limit is $\sim -0.1$ dex. We calculated binomial confidence intervals equivalent to 1$\sigma$ for the detection fractions in each mass bin following the prescription given in the Appendix of \cite{Burgasser2003}. Even with our relatively small sample, we see expected trends with stellar mass. Lower mass stars (M $< 1.3$ \msun) deplete much of their Li on the main sequence, and we find the detection of Li in the two lowest mass bins is correspondingly low. In the higher mass bins on the RGB, the detection rate is much higher. For the RGB, if we combine the two lowest mass bins and the four highest mass bins we find detection fractions of $25\%^{+19\%}_{-9\%}$ and $60\%^{+10\%}_{-13\%}$, respectively, confirming the overall difference in detectability with stellar mass.
However, one high mass bin (centered at 1.6\msun) shows a much lower fraction of Li detected stars. While this mass is not too far from the lithium dip at these metallicities  \citep{AguileraGomez2018}, analysis of the GALAH DR3 data \citep{Buder2021} indicates that the lithium dip at this metallicity should be at approximately 1.4-1.5\msun, about half a bin down from where the number of lithium detections drops (see Appendix Figure \ref{fig:galah_fig}). We therefore suggest that this could indicate a slight offset between the mass scale for the asteroseismic giants and the GALAH dwarfs, a statistical fluctuation, or that there is some sort of as yet unidentified additional lithium destruction happening on the subgiant or lower giant branch at around this mass. } 

{The detection rates of the red clump stars also shows a trend with stellar mass. In the lower mass bins, the Li detection rates are comparable to that seen in the RGB stars. However, the detection rates are much lower at higher masses. Combining the two highest mass bins we find a Li detection rate of only $14\%^{+21\%}_{-5\%}$ for the red clump compared to $75\%^{+5\%}_{-19\%}$ for the first ascent RGB.}
Such decreases in Li in the more massive stars are not inconsistent with previous results (e.g. from open clusters \citealt{Carlberg2016}, or the field \citealt{Martell2021}), but generally there have been limited samples of more massive stars from which to draw conclusions, and so more work is necessary to determine where the depletion happens and the underlying cause.
We do not note any mass bins at this metallicity where the detection rates in the clump are significantly above the rates on the red giant branch, something that might be expected if for many stars there was significant Li production and mixing during the ignition of helium burning, as predicted by some theories \citep{Casey2019, Schwab2020}. According to recent work \citep{Zhang2021, DeepakLambert2021a}, there is no empirical or theoretical indication of obvious Li depletion in the red clump phase and the Li-rich giants can be found at any point of the core-He burning evolution. Lithium in this phase should not be strongly affected by internal mixing. Thus if a giant reaches the clump with a high Li abundance it should preserve it during the clump, and we should observe it to be Li-rich. While our observations are not sufficient to rule out some complex combination of Li production, mixing, and destruction that approximately cancels itself out by the red clump, we do not see evidence for a simple enhancement of Li abundance at the tip of the red giant branch for a large fraction of stars.

{In the right panel of Figure \ref{Fig:casey}, we plot the trend of \ali\ with mass for all of the stars analyzed in this paper together with the LAMOST measurements.  The stars that were chosen based on mass and included in Figure \ref{Fig:massbins} are the filled symbols, whereas the open symbols denote stars selected by us due to their known rotation or for cross-validation. It is only among this latter sample that we find Li-rich stars. We also show the stars selected by mass separately in the right panel of Figure \ref{Fig:massbins}. This plot shows that for stars where we have Li detections, the Li abundances tend to be higher for the more massive stars, though the spread of abundances is also large.}
{In general we find that our results are consistent with the larger but less carefully constructed sample from LAMOST. In our sample, we see that there seems to be depletion in the more massive stars (Figure \ref{Fig:massbins}, left panel), but no substantial evidence for general Li creation at the tip of the red giant branch. }



\subsection{Correlations between Lithium and Rotation}

Many authors have noted a correlation between rotation rates and lithium enhanced {giants \citep[e.g.][]{FekelBalachandran1993, Drake2002, Carlberg2016, DelgadoMena2016, TakedaTajitsun2017}} {Similarly, on the main sequence and near the lithium dip, there are correlations between lithium and rotation \citep{AnthonyTwarog2021}} 
Using stars in our sample that have rotation measurements from spots, \vsini, or asteroseismology, we search for correlations between rotation and lithium abundance. 

\subsubsection{Surface Rotation}
\begin{figure}[tb]
\begin{center}

\subfigure{\includegraphics[width=0.4\textwidth,clip=true, trim=0.5in 0.2in 0.3in 0.5in]{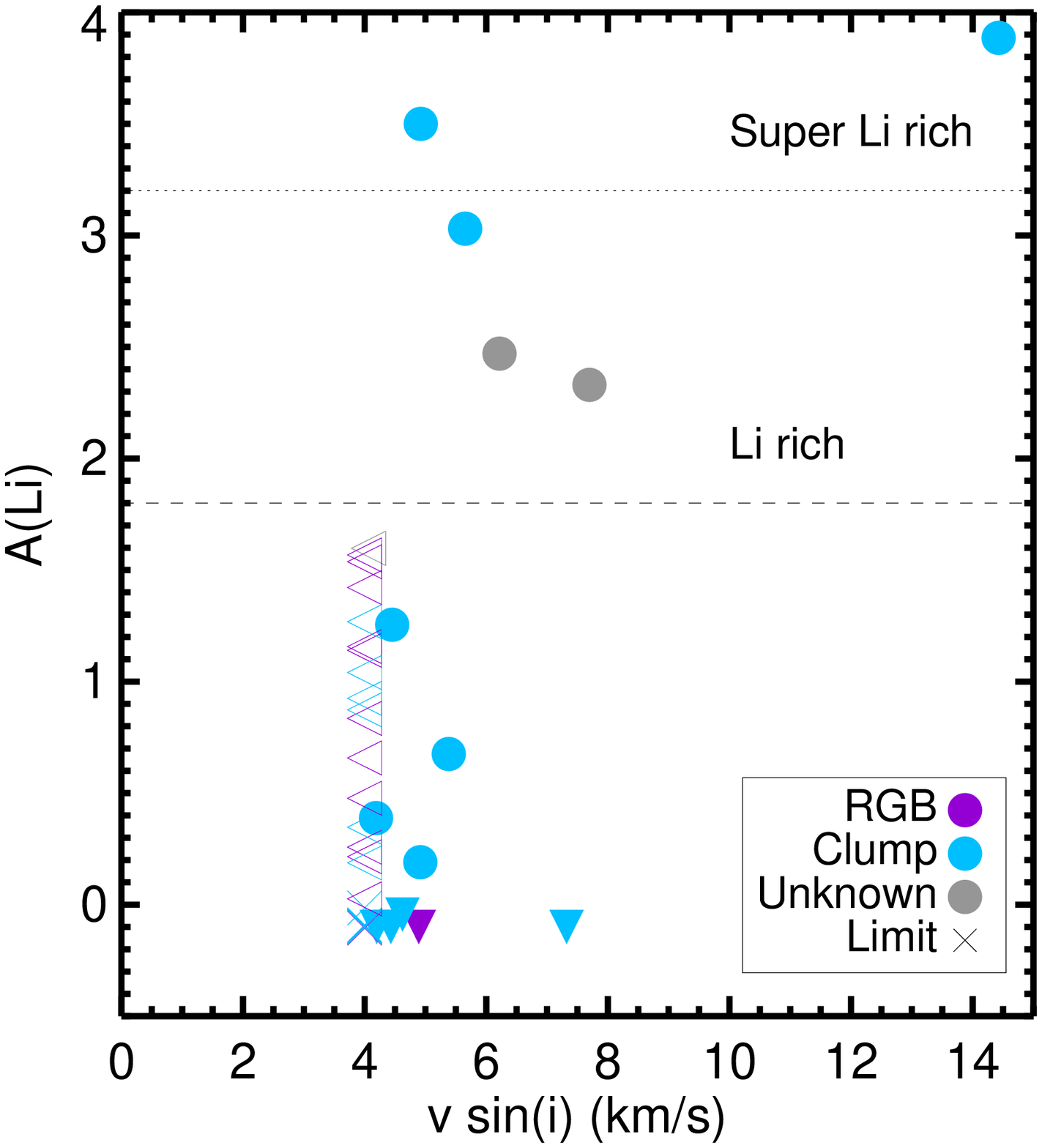}}
\subfigure{\includegraphics[width=0.4\textwidth,clip=true, trim=0.5in 0.2in 0.3in 0.5in]{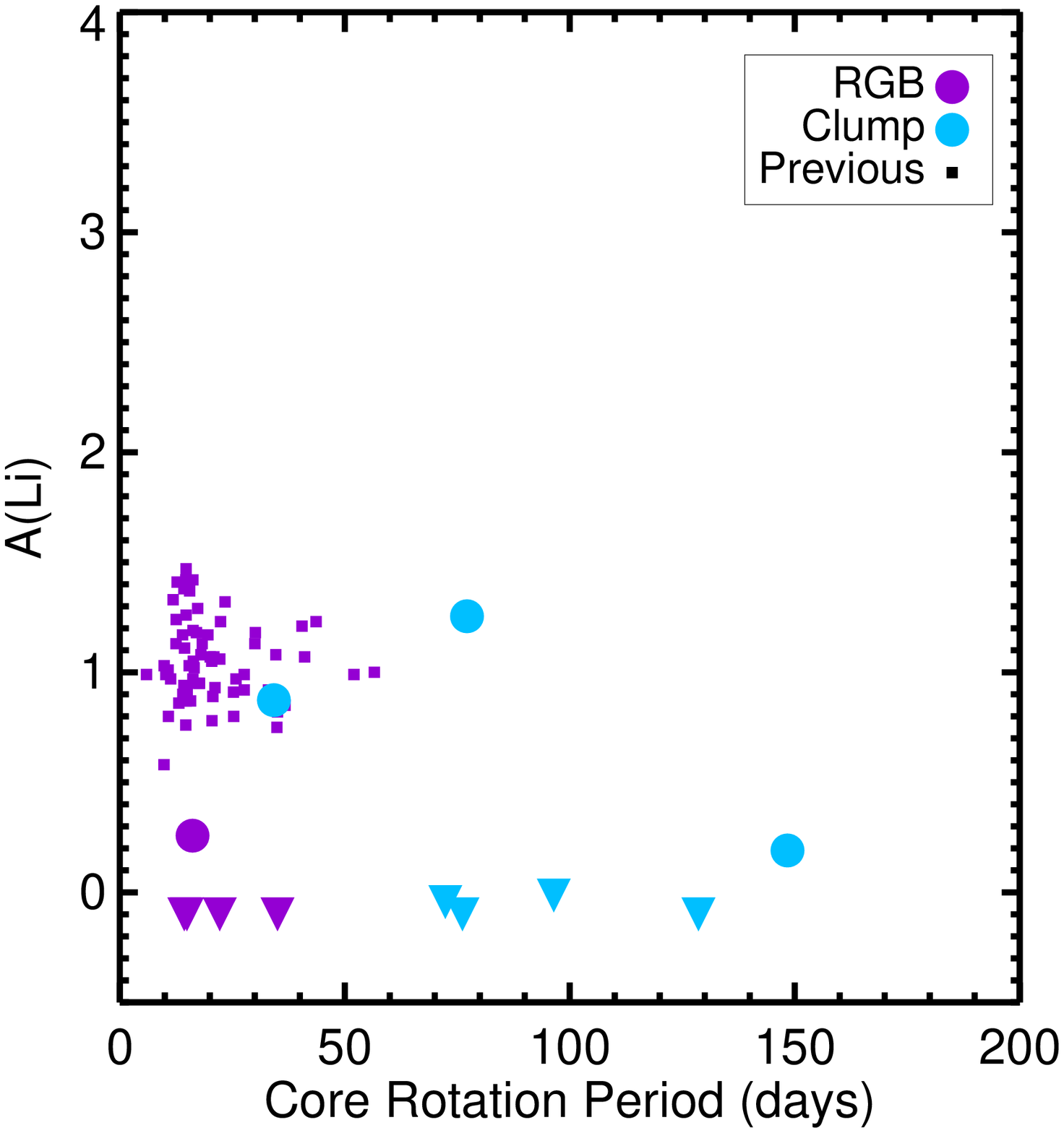}}
\caption{Lithium abundances (circles) and upper limits (downward pointing triangles) compared to rotation measurements (filled symbols) and upper limits (rightward pointing triangles). Xs represent limits in both quantities. 
\textbf{Top:} 
{We detect rotation in all of the lithium-rich stars, but there is not a strong correlation between the rotation velocity and the lithium abundance.} 
\textbf{Bottom:} No strong correlations are seen between the core rotation period and Li abundance, although stars with strong Li enhancement are less likely to have measured core rotation periods, possibly due to complications in seismology of rapidly rotating stars. Previous data from LAMOST \citep{Gao2021} are shown as small purple squares. } 

\label{Fig:rotation-lithium}
\end{center}
\end{figure}

In general, \citep[see e.g.][]{Massarotti2008,Tayar2015,Ceillier2017, TayarPinsonneault2018}, the rotation rates of giant stars are expected to be slow, with low velocities {($<3$ \kms)} and long periods {(hundreds to thousands of days)}. However, 
as part of our sample selection, we included some stars known to rotate rapidly either from their spectroscopic rotation velocities \citep{Tayar2015} or from their spot rotation periods {\citep{Ceillier2017}}. In the interim, we have added to this sample any stars that have rotation periods quoted in \citet{Gaulme2020}, rotation velocities quoted in \citet{Daher2022}, as well as  
\vsini measurements or limits from the Subaru spectra used in this study. {Most of the stars with detectable rotation from any method had measured rotation velocities from the Subaru spectra, and so we have generally plotted those values. However, in a few cases the rotation velocities were near or below our detection limit, but we were able to convert the rotation period to a velocity using the asteroseismic radii to plot those points as rotation detections.}
For stars that are not spotted, and whose rotation velocities are too slow to measure, we assume an upper limit on the rotation velocity of 
4 \kms\ from the Subaru {spectra.} In Figure \ref{Fig:rotation-lithium}, we show all of our estimates of surface rotation, compared to our Li measurements (circles) and limits (downward pointing triangles). 
{We do not see a simple }correlation between Li abundance and rotation period in our data, but we do not have very many stars in the super-Li-rich regime where such correlations have been claimed \citep{Du2021}.
Consistent with previous authors, {we find that lithium-rich giants tend to be rapidly rotating -- in our case we detect measureable rotation in all five lithium-rich stars. Four of these were included in this work for their previously known high lithium abundance. However, high rotation did not guarantee lithium-richness. Of the nine stars with detected rotation and no previous lithium measurement, only one was found to be lithium-rich. This is still a higher rate than the much larger population of unmeasurably slow rotators, among which we find zero Li-rich stars.} 
Since find many Li-poor stars among the faster rotators and Li detections among the slower rotators, { we suggest that additional data would be required to better study the relationship between lithium and rotation, as the relationship between the two is not simple.} 

{In general, we expect that any of the stars with measurable rotation in our sample are rapidly rotating because they have gained angular momentum on the red giant branch through an interaction with a stellar or substellar companion. While some of the stars in are sample are 
more massive than the Kraft break \citep{Kraft1967}, and in theory their rotation could be retained from their rapid rotation on the main sequence, in practice, many authors \citep{Massarotti2008,Tayar2015,Deheuvels2015, Carlberg2016, Ceillier2017} have found that, in practice, these more massive stars rotate more slowly than expected in the core-helium-burning phase, likely as a result of enhanced angular momentum loss \citep{TayarPinsonneault2018}. We therefore argue that their detected rotation in our analysis is unlikely to be the result of angular momentum retained from the main sequence, and much more likely to be the result of an interaction.} 




\subsubsection{Core Rotation}


\begin{figure}[htb]
\begin{center}

{\includegraphics[width=0.5\textwidth,clip=true, trim=0.5in 0in 0in 0in]{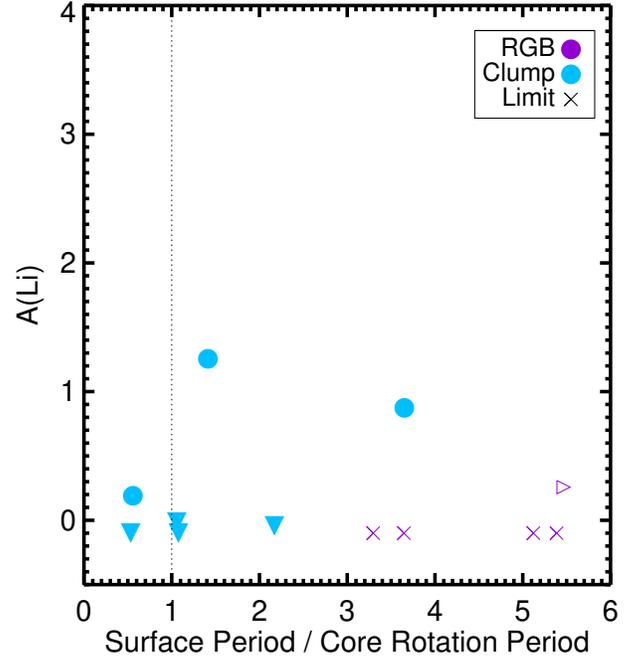}}

\caption{Lithium abundances (circles) and upper limits (downward pointing triangles) compared to the ratio (filled symbols) or limits (rightward triangles or Xs) of the surface rotation period to the core rotation period. As in previous plots, the evolutionary states of the points are indicated by the color (purple for RGB, blue for clump, grey for ambiguous). Stars rotating as a solid body, and therefore presumably with minimal shear forces to drive mixing, would have a ratio of 1 (dotted vertical line). We see no correlation between the core-surface contrast, and therefore the presumed shear forces and rotational mixing, and the detection or abundance of lithium at the stellar surface. } 

\label{Fig:rotationgradient}
\end{center}
\end{figure}

In some theories \citep[e.g.][]{Zahn1992} it is the internal rotation profile, rather than the surface rate, that should dictate the rotational mixing in the stellar interior and thus the surface abundance of Li. In the past, there was no way of estimating the interior rate, and so it had to be assumed to scale with the surface rotation rate. However, in the past decade it has become possible to infer the rotation rates of the cores of giant stars from their oscillation spectra due to the mixing of the gravity modes in the core with the pressure modes observed at the surface \citep{Beck2011, Deheuvels2012, Mosser2012b}. In our sample, we specifically targeted a set of stars with core and surface rotation rates available from \citet{Tayar2019b}. In the course of our analysis, we realized that several stars in our sample also had core rotation measurements available from \citet{Gehan2018}. In the lower panel of Figure \ref{Fig:rotation-lithium}, we show the Li measurements and limits compared to the inferred core rotation period for our stars. We also show on this plot all stars with core rotation measurements available from \citet{Gehan2018} and Li abundances available from LAMOST \citep{Gao2021} that were in the APOKASC sample; in the interest of sample size for comparison, we do not apply any cuts in effective temperature or metallicity. In the combination of the two data sets, we find no correlation between the core rotation rate and the measurement or abundance of Li at the surface. We do however note that detailed asteroseismology, including the estimation of the interior rotation rate, can be more challenging in active and rapidly rotating stars (\citealt{Gaulme2014}, but see also \citealt{Mathur2020}), which might bias core rotation detections against the most Li rich and rapidly rotating stars.

There have also been suggestions \citep{ TalonCharbonnel2003, denissenkov2009} that the local shear forces caused by rotational {{gradients}} are important for driving mixing. We therefore show in Figure \ref{Fig:rotationgradient} the ratio of the surface rotation period to the core rotation period, as an estimate of the total internal shears. Stars with a ratio of 1 are consistent with rotating as solid bodies and therefore presumably have minimal shear forces. Most stars have surfaces rotating more slowly than their cores, consistent with the expectations of single star evolution \citep{TayarPinsonneault2018, Tayar2019b}, while there are a few stars whose measurements suggest that their surfaces are rotating faster than their cores (ratios less than 1). While some of these could be measurement errors \citep{Tayar2019b}, there are some stars where such rotation profiles seem to be present \citep{Kurtz2014,Tayar2022c} and they are generally explained with angular momentum transfer from interaction with a companion \citep[e.g.][]{Daher2022}. Should our stars with surfaces rotating faster than their cores prove to be robust, they would be in conflict with theories like that presented in \citet{Casey2019}, which suggest that tidal interactions drive Li enhancements that should persist longer than the resulting rotation.
More generally, when we look at our stars that have both core and surface rotation estimates or limits and Li abundances, we do not see any strong correlations that would suggest a significant impact of shear mixing on the Li abundances of giants. 


\subsection{Correlations with Indications of Binarity}
\begin{figure}[tb]
\begin{center}

\includegraphics[width=0.4\textwidth,clip=true, trim=0.5in 0.2in 0.3in 0.5in]{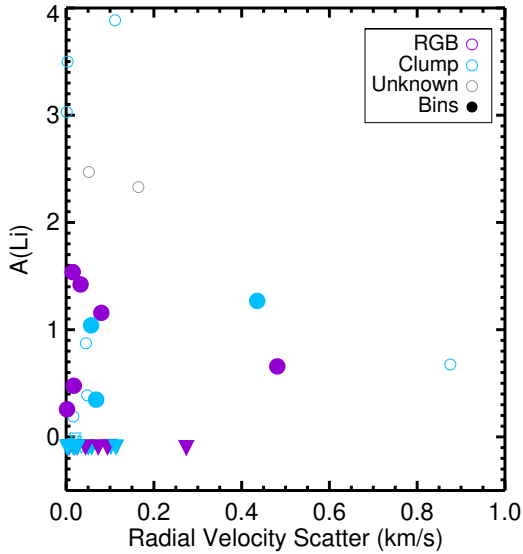}
\caption{ 
{No strong relationships are seen between the radial velocity scatter, a coarse indicator of close binaries, and the lithium abundance, although there is perhaps a slight tendency for stars selected in mass bins to be more likely to have higher radial velocity scatter if they have measured lithium.}} 

\label{Fig:binarityRV}
\end{center}
\end{figure}

Many authors have suggested that Li richness, either on the red giant branch {\citep{Carlberg2016,AguileraGomez2016, DelgadoMena2016, Soares-Furtado2021}} or in the red clump \citep{Casey2019} should be related to the interaction of a giant with a stellar or substellar companion. Because our stars are so exquisitely characterized, we can look at a variety of properties that correlate with binarity and see if they have any correlations with the Li abundance. Because the APOGEE survey used fixed observing times, the stars in the APOKASC sample were often observed multiple times to build up sufficient signal-to-noise; as part of that process APOGEE also makes available measurements of the radial velocity scatter between observations. We show {in Figure \ref{Fig:binarityRV}} that while there are a few stars that show evidence of radial velocity variability from a close companion with measured Li, there is no particularly strong correlation between radial velocity scatter and Li abundance, and not all Li-rich stars show evidence for significant radial velocity scatter. 

Gaia DR3 \citep{GaiaDR3binary} provides some information about the binary nature of our target stars through the flag NON\_SINGLE\_STAR, indicating astrometric, spectroscopic, or eclipsing binaries. Only 9 of the 63 giants in the sample are considered non single, all of them with A(Li) $<1.2$ dex. However, the Gaia selection of binaries is not complete, and thus it is not possible with this information alone to discard a possible relation between lithium enhancement and the presence of binary companions (e.g., \cite{sayeed2023}, M. Castro-Tapia et al. 2023, in prep).

\begin{figure}[tb]
\begin{center}

\includegraphics[width=0.4\textwidth,clip=true, trim=0.5in 0.2in 0.3in 0.5in]{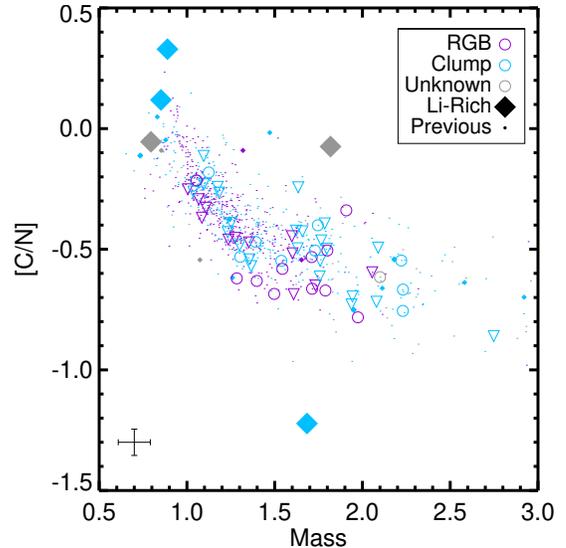}
\caption{ 
 { The [C/N] ratio is a mixing diagnostic that correlates with mass for both clump (blue) and RGB (purple) stars. Most sample stars with lithium abundances (circles) and upper limits (downward pointing triangles) follow this correlation. However, stars that are lithium-rich (diamonds) seem to deviate from the relation; deviations tend to be more common in binary evolution products.}} 

\label{Fig:binarityCN}
\end{center}
\end{figure}

One other way of identifying stars that are undergoing or have undergone interactions is through their chemistry. On the red giant branch, we generally expect the mass of a star to correlate with its carbon to nitrogen ratio [C/N] resulting from the mass dependence of the first dredge up, which has been useful for a variety of galactic archaeology purposes \citep[][J. Roberts, 2023 in prep]{Martig2016, Ness2016}. We show {in Figure \ref{Fig:binarityCN}} that most of the stars in our sample follow this correlation with perhaps slight differences in the relationship for the clump and first ascent giant stars. However, we mark as larger diamonds the Li-rich stars in our sample, and note that all five of them seem to be offset from the general population. We note that this offset for Li-rich giants is not quite as clear in the larger sample from \citet{Gao2021} (shown as smaller background points), and so we encourage further exploration of this point. If it turns out that there is a significant subpopulation of Li-rich red giants with [C/N] ratios that do not match their current masses, it may be a good tracer for stars that have undergone significant mass transfer, and thus their mass during the first dredge up was not the same as their current mass, or indicate that whatever mixing process is impacting the Li abundance is also reaching deeper into the interior, where the [C/N] ratio is set. However, it is also possible that increased rotation leads to poorer spectroscopic fits {\citep{Patton2023}} and mis-measured abundances, which can incidentally push stars off of the normal relationship.


Mass transfer or past interaction with a binary companion has also been invoked to explain a different type of unusual objects, young alpha-rich stars \citep{Martig2015}. Although these stars are not thought to be directly related to the phenomenology of Li-rich giants, and our sample giants are intrinsically more metal-rich than most of the galactic alpha-rich population, the young alpha-rich stars also show an unusual behavior in the [C/N]-mass relation \citep{Jofre2022}, with most of them located outside the general population trend, suggesting that offsets in mixing sensitive ratios like the ratio of carbon to nitrogen {(E. Bufanda et al., submitted)} or the ratio of carbon-12 to carbon-13 \citep{AguileraGomez2022} may be diagnostic of a wide variety of binary interaction processes. 



\section{Discussion} \label{Sec:Discussion}

In this work, we have looked at an extremely well-characterized sample of metal-rich red giants in order to get a clearer picture of the distribution of Li in these stars. Previous work has suggested a complex set of processes are impacting Li in red giants, and our analysis seems to agree with that view. {Our results are consistent with previous work that super Li rich red giants are rare, that Li-rich giants are more common in the core-helium-burning phase, that rapid surface rotation is often associated with higher Li abundances, but there is no simple relationship. We are also} able to demonstrate that:

\begin{itemize}
    \setlength\itemsep{0em}
    \setlength\parskip{0em}
    \item LAMOST low and medium resolution spectra are entirely sufficient to accurately identify Li rich giants
    \item the baseline Li abundance of red giants is likely mass-dependent with more massive red giants generally having higher Li
    \item in stars more massive than $\sim$ 1.8 \msun, Li destruction or depletion is likely happening on the upper red giant branch
    \item core rotation and the core-surface rotational shear seem to be uncorrelated with Li abundance 
    \item Li-rich giants may also have offsets in their carbon to nitrogen ratios
    \item binarity may be related to the Li rich phenomenon in some cases, but it is likely neither necessary nor sufficient
\end{itemize}


Given how our targets were selected, we are not able to determine a fraction of enriched RGB or red clump stars. To do this, it is not only necessary to have a large sample of stars, but also, to define first what is considered a truly lithium-rich giant based on the abundances of other stars of similar mass, metallicity, and evolutionary stage. The more massive RGB stars could naturally give rise to more massive enriched red clump stars, while different mechanisms could be acting to produce enrichment in the RGB or less massive red clump giants. Regardless of the specific mechanism, stellar mass is key, and other indicators, such as the carbon-to-nitrogen ratio may provide the additional information needed to distinguish between processes \citep[see e.g.,][] {Zhou2022}.

The Li-rich giants continue to be one of the most interesting and frustrating questions in stellar physics. It is clear that these objects have interesting stories to tell about stellar histories that include information about binarity, rotation, mass, metallicity, mixing, and possibly planets, but teasing out the details of those stories has continued to prove challenging. As the number of Li measurements continues to increase, and the complementary knowledge including stellar masses, ages, evolutionary states, binary companions, and so forth becomes more common and precise, we can only hope that eventually some physical explanation, or more likely some combination of physical explanations, will be able to identify the reason for Li enrichment in both a population sense, and on a star-by-star basis.

\begin{acknowledgements}

We thank the referee for helpful suggestions that improved this manuscript. We thank Travis Berger for his help with the preparation of the Subaru observation files. We thank the Maunakea and Subaru staff for their assistance with these observations.
This research is based on data collected at the Subaru Telescope, which is operated by the National Astronomical Observatory of Japan. We are honored and grateful for the opportunity of observing the Universe from Maunakea, which has the cultural, historical, and natural significance in Hawaii.
The authors wish to recognize and acknowledge the very significant cultural role and reverence that the summit of Maunakea has always had within the indigenous Hawaiian community.  We are most fortunate to have the opportunity to conduct observations from this mountain.

\end{acknowledgements}

\begin{longrotatetable}
\begin{deluxetable*}{rlrrrrrrrrrrrrrrr}
\tablecaption{Stellar properties for all stars in our sample.\label{table1}}
\tablewidth{700pt}
\tabletypesize{\scriptsize}
\label{Tab:table1}
\tablehead{\colhead{KICID} &           \colhead{2MASS ID} &             \colhead{Gaia DR2 ID} &  \colhead{Mass} &  \colhead{Log(g)} &   \colhead{[Fe/H]} &  \colhead{[$\alpha$/Fe]} & \colhead{T$_\mathrm{eff}$} &\colhead{[C/N]} & \colhead{$\textrm{A(Li)}$} & \colhead{$\textrm{A(Li)}_{UL}$} &  \colhead{Flag} &  \colhead{$v\sin(i)$} &  \colhead{Period$_\textrm{s}$} &  \colhead{Period$_\textrm{c}$} &   \colhead{$\sigma_v$} & \colhead{State} \\
\colhead{} &           \colhead{} &             \colhead{} &  \colhead{\msun} &  \colhead{dex} &   \colhead{dex} &  \colhead{dex} & \colhead{K} &\colhead{dex} & \colhead{dex} & \colhead{} &  \colhead{} &  \colhead{\kms} &  \colhead{days} &  \colhead{days} &   \colhead{\kms} & \colhead{}} 
\startdata
10461323 & 2M19114879+4741524 & 2130879480532813824 & 2.10000 & 2.86000 &   0.118 &      -0.054 & 4874.06 & -0.616 &   1.60 &         0 &     3 &  4.065 &              79.7 &        -9999.0 &   -9999.000 &      U \\
10522084 & 2M19024774+4745344 & 2131483245555109632 & 1.62200 & 2.57600 &   0.043 &       0.006 & 4833.55 & -0.423 &  -0.09 &         1 &     0 &  3.264 &           -9999.0 &        -9999.0 &   -9999.000 &     RC \\
10550429 & 2M19492290+4746588 & 2086419800156327680 & 1.49700 & 3.03900 &   0.087 &      -0.031 & 4872.35 & -0.685 &   1.16 &         0 &     0 &  2.885 &           -9999.0 &        -9999.0 &       0.080 &    RGB \\
10587122 & 2M19033449+4748388 & 2131504857830540032 & 1.23200 & 2.48200 &   0.187 &      -0.011 & 4700.81 & -0.429 &  -0.10 &         1 &     0 &  3.529 &           -9999.0 &        -9999.0 &   -9999.000 &     RC \\
10722175 & 2M19122333+4803442 & 2130984827490239232 & 1.35300 & 2.95700 &   0.232 &      -0.028 & 4736.14 & -0.472 &  -0.10 &         1 &     0 &  2.620 &           -9999.0 &           22.2 &   -9999.000 &    RGB \\
10793771 & 2M19213869+4809014 & 2129414106407732608 & 2.05500 & 2.80600 &   0.139 &      -0.012 & 4873.17 & -0.596 &  -0.10 &         1 &     0 &  3.385 &           -9999.0 &        -9999.0 &   -9999.000 &    RGB \\
11087371 & 2M19335280+4836440 & 2129006870493887744 & 1.28400 & 3.06100 &   0.296 &       0.013 & 4750.13 & -0.621 &   0.21 &         0 &     0 &  3.731 &           -9999.0 &            inf &   -9999.000 &    RGB \\
11358669 & 2M19425925+4911303 & 2134772129653795328 & 1.39800 & 2.89600 &   0.183 &       0.008 & 4755.77 & -0.631 &   0.26 &         0 &     0 &  3.514 &           -9999.0 &           16.2 &       0.002 &    RGB \\
11496569 & 2M19025133+4929261 & 2132105435996154368 & 1.59800 & 2.73300 &   0.100 &      -0.007 & 4786.56 & -0.446 &  -0.10 &         1 &     0 &  3.272 &           -9999.0 &        -9999.0 &   -9999.000 &    RGB \\
11550492 & 2M19065989+4932243 & 2131326599508772352 & 1.10800 & 2.87800 &   0.055 &       0.003 & 4737.56 & -0.327 &  -0.10 &         1 &     0 &  3.218 &           -9999.0 &           14.9 &   -9999.000 &    RGB \\
11615224 & 2M19363394+4938350 & 2135022302904783360 & 0.85200 & 2.38200 &   0.015 &       0.024 & 4700.81 &  0.119 &   3.03 &         0 &     2 &  5.653 &           -9999.0 &        -9999.0 &       0.001 &     RC \\
 1161618 & 2M19242614+3648478 & 2050254968627140736 & 1.18300 & 2.43800 &   0.064 &       0.005 & 4741.98 & -0.265 &  -0.10 &         1 &     3 &  4.199 &             158.3 &        -9999.0 &       0.021 &     RC \\
11618522 & 2M19421336+4939595 & 2134930768564971776 & 1.94300 & 2.67400 &   0.189 &      -0.020 & 4865.36 & -0.695 &  -0.09 &         1 &     0 &  3.349 &           -9999.0 &        -9999.0 &       0.102 &    2CL \\
11649294 & 2M18562652+4945381 & 2132278815233502720 & 1.09600 & 2.43000 &   0.100 &       0.004 & 4753.13 & -0.231 &  -0.10 &         1 &     0 &  3.337 &           -9999.0 &        -9999.0 &   -9999.000 &     RC \\
11969378 & 2M19390565+5019540 & 2135158951586941184 & 1.35600 & 2.49100 &   0.090 &       0.002 & 4752.94 & -0.542 &  -0.10 &         1 &     0 &  3.059 &           -9999.0 &        -9999.0 &       0.050 &     RC \\
12070114 & 2M19430023+5032393 & 2135361394864817152 & 1.63400 & 2.50700 &   0.130 &       0.020 & 4756.96 & -0.243 &  -0.10 &         1 &     4 &  3.367 &           -9999.0 &        -9999.0 &       0.035 &     RC \\
 2015820 & 2M19265758+3728035 & 2051784840284088064 & 1.79700 & 2.56500 &   0.037 &      -0.009 & 4827.03 & -0.498 &  -0.09 &         1 &     0 &  3.816 &           -9999.0 &        -9999.0 &       0.015 &     RC \\
 2305930 & 2M19282563+3741232 & 2051887163583939200 & 0.88900 & 2.36900 &  -0.435 &       0.165 & 4858.64 &  0.329 &   3.89 &         0 &     2 & 14.427 &              33.0 &        -9999.0 &       0.111 &     RC \\
 3339894 & 2M19232223+3826312 & 2052834599006426240 & 1.81800 & 2.38900 &  -0.223 &       0.037 & 4748.55 & -0.074 &   2.33 &         0 &     2 &  7.698 &             100.0 &        -9999.0 &       0.165 &      U \\
 3459109 & 2M19404372+3830503 & 2049134982905799552 & 1.78900 & 2.69600 &   0.036 &      -0.006 & 4832.87 & -0.671 &   1.54 &         0 &     0 &  3.424 &           -9999.0 &        -9999.0 &       0.015 &    RGB \\
 3526061 & 2M19004471+3836399 & 2100133252769956224 & 1.27400 & 3.01700 &   0.224 &       0.002 & 4770.27 & -0.453 &  -0.10 &         1 &     0 &  2.775 &           -9999.0 &           14.3 &   -9999.000 &    RGB \\
 3526625 & 2M19014365+3840029 & 2100131637861896192 & 1.12400 & 2.44400 &   0.108 &       0.029 & 4730.83 & -0.182 &   0.39 &         0 &     3 &  4.187 &             125.1 &        -9999.0 &       0.048 &     RC \\
 3860253 & 2M19360616+3854028 & 2052378026807507968 & 1.71100 & 2.84500 &   0.000 &      -0.024 & 4867.49 & -0.532 &   1.42 &         0 &     0 &  3.193 &           -9999.0 &        -9999.0 &       0.033 &    RGB \\
 3958400 & 2M19301845+3904562 & 2053093396558506112 & 2.08900 & 2.64500 &   0.139 &      -0.009 & 4878.66 & -0.493 &  -0.10 &         1 &     0 &  3.302 &           -9999.0 &        -9999.0 &       0.018 &    2CL \\
 4055294 & 2M19241086+3907357 & 2052965234733414528 & 1.60800 & 2.99300 &   0.183 &       0.009 & 4861.87 & -0.686 &  -0.11 &         1 &     0 &  2.796 &           -9999.0 &        -9999.0 &       0.274 &    RGB \\
 4474383 & 2M19370790+3934123 & 2052437396129089664 & 1.93700 & 2.60100 &   0.054 &      -0.013 & 4894.66 & -0.728 &  -0.10 &         1 &     0 &  3.696 &           -9999.0 &        -9999.0 &       0.019 &     RC \\
 4633907 & 2M18564805+3946115 & 2103296925679588992 & 1.09500 & 2.38700 &   0.049 &       0.027 & 4704.50 & -0.113 &  -0.10 &         1 &     0 &  4.431 &           -9999.0 &        -9999.0 &   -9999.000 &     RC \\
 4648485 & 2M19193317+3945381 & 2101044507393877248 & 1.06700 & 3.05700 &   0.093 &       0.049 & 4726.61 & -0.293 &  -0.10 &         1 &     0 &  2.299 &           -9999.0 &        -9999.0 &       0.044 &    RGB \\
 4826087 & 2M19154014+3957514 & 2101241629211118720 & 1.60100 & 2.57300 &   0.037 &       0.007 & 4714.82 & -0.517 &  -0.10 &         1 &     0 &  3.333 &              16.0 &        -9999.0 &       0.007 &    RGB \\
 4913049 & 2M19135948+4005110 & 2101299211835208704 & 1.00500 & 3.24800 &   0.005 &       0.092 & 4777.97 & -0.251 &  -0.10 &         1 &     0 &  4.891 &           -9999.0 &        -9999.0 &       0.073 &    RGB \\
 4931389 & 2M19353795+4004428 & 2052513438539100544 & 1.23600 & 3.22500 &   0.142 &      -0.016 & 4838.31 & -0.460 &  -0.10 &         1 &     0 &  2.909 &           -9999.0 &        -9999.0 &       0.025 &    RGB \\
 4937056 & 2M19411631+4005508 & 2076298383311466368 & 1.78500 & 2.58800 &   0.101 &      -0.031 & 4850.22 & -0.393 &  -0.10 &         1 &     5 &  7.322 &              82.0 &           76.2 &       0.026 &     RC \\
 5087190 & 2M19115123+4013021 & 2100556800967018368 & 2.08000 & 2.74700 &   0.250 &      -0.021 & 4904.59 & -0.718 &  -0.04 &         1 &     5 &  4.623 &             157.0 &           72.3 &       0.023 &    2CL \\
 5095946 & 2M19230745+4017484 & 2101132159082055552 & 1.25200 & 2.46600 &   0.140 &      -0.003 & 4760.51 & -0.391 &  -0.10 &         1 &     0 &  3.813 &           -9999.0 &        -9999.0 &       0.005 &     RC \\
 5194072 & 2M19352306+4023194 & 2077289970993761408 & 2.23075 & 2.63582 &   0.285 &      -0.002 & 4777.94 & -0.667 &   0.19 &         0 &     5 &  4.916 &              82.7 &          148.4 &       0.016 &  -9999 \\
 5383359 & 2M19482276+4032018 & 2073730439545220224 & 1.05100 & 2.42100 &   0.035 &       0.047 & 4734.59 & -0.215 &   0.35 &         0 &     0 &  3.335 &           -9999.0 &        -9999.0 &       0.068 &     RC \\
 5534910 & 2M19285814+4046073 & 2053573466520267520 & 2.74800 & 2.95800 &   0.279 &      -0.007 & 4949.62 & -0.860 &  -0.01 &         1 &     5 &  3.422 &             102.0 &           96.4 &       0.021 &    2CL \\
 5723909 & 2M19465321+4058004 & 2076757223253430144 & 1.53800 & 2.51700 &   0.243 &      -0.029 & 4728.04 & -0.548 &   0.92 &         0 &     0 &  3.033 &           -9999.0 &        -9999.0 &   -9999.000 &     RC \\
 5769244 & 2M18534059+4102118 & 2103697147911515136 & 1.08500 & 3.04800 &   0.240 &       0.019 & 4719.54 & -0.369 &  -0.10 &         1 &     0 &  2.335 &           -9999.0 &        -9999.0 &   -9999.000 &    RGB \\
 5791889 & 2M19283210+4104584 & 2053597934956366464 & 1.71300 & 3.00400 &   0.197 &      -0.020 & 4887.23 & -0.664 &   0.84 &         0 &     0 &  3.131 &           -9999.0 &        -9999.0 &   -9999.000 &    RGB \\
 5811766 & 2M19490669+4101352 & 2076770039436573952 & 1.80000 & 2.55600 &   0.082 &      -0.003 & 4720.23 & -0.505 &   1.57 &         0 &     0 &  3.158 &           -9999.0 &        -9999.0 &   -9999.000 &    RGB \\
 5943345 & 2M19015661+4114403 & 2103856267861037824 & 2.22100 & 2.61100 &   0.102 &      -0.010 & 4895.35 & -0.547 &   1.25 &         0 &     5 &  4.452 &             109.0 &           77.2 &   -9999.000 & RC/2CL \\
 6032639 & 2M19141495+4118432 & 2102124743207232128 & 1.74600 & 2.56800 &   0.018 &      -0.023 & 4861.82 & -0.401 &   0.68 &         0 &     3 &  5.384 &             130.0 &        -9999.0 &       0.876 &     RC \\
 6103934 & 2M18564594+4128038 & 2103725284237785728 & 1.54400 & 2.90600 &   0.120 &      -0.034 & 4824.73 & -0.581 &   1.14 &         0 &     0 &  3.872 &           -9999.0 &        -9999.0 &   -9999.000 &    RGB \\
 7017044 & 2M19031889+4230179 & 2104024012104377344 & 1.03600 & 2.40900 &   0.016 &       0.022 & 4746.53 & -0.261 &  -0.10 &         1 &     0 &  3.543 &           -9999.0 &        -9999.0 &       0.006 &     RC \\
 7097136 & 2M18562647+4237349 & 2104920492036763520 & 1.63100 & 2.51700 &   0.133 &      -0.010 & 4724.21 & -0.496 &  -0.10 &         1 &     0 &  3.427 &           -9999.0 &        -9999.0 &   -9999.000 &     RC \\
 7098510 & 2M18585908+4237588 & 2104250786376888448 & 1.30000 & 2.48000 &   0.071 &      -0.006 & 4787.58 & -0.484 &  -0.10 &         1 &     0 &  3.599 &           -9999.0 &        -9999.0 &   -9999.000 &     RC \\
 7283405 & 2M19285388+4250513 & 2125813583790089088 & 1.73300 & 2.75600 &   0.005 &      -0.002 & 4857.25 & -0.650 &  -0.10 &         1 &     0 &  3.070 &           -9999.0 &           35.0 &       0.094 &    RGB \\
 7351098 & 2M19114520+4255162 & 2102592482324682368 & 2.23000 & 2.97500 &   0.315 &      -0.003 & 4916.38 & -0.756 &   0.87 &         0 &     5 &  3.292 &             125.0 &           34.2 &       0.045 &    2CL \\
 7457184 & 2M19410681+4303056 & 2078012762449199360 & 1.97400 & 2.73600 &   0.342 &       0.007 & 4778.26 & -0.782 &   0.03 &         0 &     0 &  2.627 &           -9999.0 &        -9999.0 &   -9999.000 &    RGB \\
 7499531 & 2M18414542+4308496 & 2116826072661485312 & 1.90800 & 2.75200 &   0.091 &       0.080 & 4821.27 & -0.339 &   0.66 &         0 &     0 &  3.223 &           -9999.0 &        -9999.0 &       0.481 &    RGB \\
 7585122 & 2M18500383+4316049 & 2105182004005398656 & 1.36600 & 2.50100 &   0.239 &      -0.004 & 4705.34 & -0.570 &  -0.10 &         1 &     0 &  3.159 &           -9999.0 &        -9999.0 &       0.114 &     RC \\
 8197210 & 2M20033477+4402323 & 2075875170103511040 & 1.76500 & 2.56700 &   0.035 &      -0.027 & 4867.05 & -0.463 &  -0.10 &         1 &     0 &  3.309 &           -9999.0 &        -9999.0 &       0.002 &     RC \\
 8540767 & 2M18502221+4436052 & 2105409499833058048 & 0.79500 & 2.36400 &  -0.149 &       0.026 & 4883.28 & -0.055 &   2.47 &         0 &     2 &  6.218 &           -9999.0 &        -9999.0 &       0.052 &      U \\
 8555914 & 2M19210796+4437273 & 2126987037571830912 & 1.17500 & 2.45800 &   0.050 &      -0.020 & 4719.37 & -0.242 &  -0.10 &         1 &     0 &  3.164 &           -9999.0 &        -9999.0 &       0.018 &     RC \\
 8782196 & 2M20023887+4455347 & 2082151839616048768 & 1.05600 & 2.72500 &   0.011 &       0.034 & 4701.62 & -0.215 &   0.48 &         0 &     0 &  2.733 &           -9999.0 &        -9999.0 &       0.017 &    RGB \\
 8872709 & 2M19053513+4506383 & 2106419538702242304 & 1.39200 & 2.49100 &   0.016 &       0.015 & 4838.61 & -0.470 &   1.04 &         0 &     0 &  3.138 &           -9999.0 &        -9999.0 &       0.057 &     RC \\
 8879518 & 2M19181645+4506527 & 2127061804363014272 & 1.68400 & 2.57200 &   0.163 &      -0.004 & 4839.97 & -1.222 &   3.50 &         0 &     3 &  4.924 &             109.0 &        -9999.0 &       0.003 &     RC \\
 9335570 & 2M19151622+4553087 & 2127334551963243136 & 1.30400 & 2.47100 &   0.001 &      -0.001 & 4811.13 & -0.532 &   0.19 &         0 &     0 &  3.343 &           -9999.0 &        -9999.0 &   -9999.000 &     RC \\
 9469212 & 2M19341942+4604596 & 2128063768695712256 & 1.75600 & 2.54200 &   0.297 &      -0.013 & 4722.75 & -0.613 &  -0.10 &         1 &     5 &  3.907 &              68.4 &          128.6 &   -9999.000 &     RC \\
 9491316 & 2M20013827+4604468 & 2085315233343597696 & 1.72900 & 2.63100 &   0.111 &      -0.020 & 4856.02 & -0.505 &   1.27 &         0 &     0 &  3.009 &           -9999.0 &        -9999.0 &       0.436 & RC/2CL \\
 9575645 & 2M19022297+4615098 & 2106548314702107264 & 1.65800 & 2.57100 &   0.113 &       0.006 & 4841.36 & -0.426 &  -0.10 &         1 &     0 &  3.201 &           -9999.0 &        -9999.0 &       0.026 &     RC \\
 \enddata
\tablecomments{Lithium values quoted here include NLTE corrections. 
Mass refers to the empirically corrected version of the seismic scaling relation mass, equivalent to the APOKASC2 seismic mass from \cite{Pinsonneault2018}. Similarly, $\log g$ refers to the APOKASC2 seismic $\log g$. {[Fe/H], [$\alpha$/Fe], T$_{\rm eff}$, and [C/N] values are taken from APOGEE DR14.} $\sigma_v$ refers to the radial velocity scatter from APOGEE DR14. {Lithium upper limits are marked as 1 in the A(Li)$_{UL}$ column.}  Period$_s$ refers to the best estimate for the surface rotation period from \cite{Ceillier2017} or \cite{Gaulme2020}. Period$_c$ refers to the core rotation period from \cite{Tayar2019b} or \cite{Gehan2018}. Stars without measured surface or core rotation periods are allocated a period of -9999. Flag indicates the reason for selection, 0: Mass bins, 2: lithium rich, 3: spots, 4: velocity, 5: \citet{Tayar2019b}. {Evolutionary states are taken from \citet{Pinsonneault2018} and described further in \citet{Elsworth2019}, U: Uncertain, RGB: red giant branch, RC: red clump, 2CL: secondary clump, -9999: no answer was returned} } 
\end{deluxetable*}
\end{longrotatetable}

 \restartappendixnumbering
\appendix
\section{Lithium Dip}

\begin{figure}[h]
    \centering
    \includegraphics[width=0.5\textwidth,clip=true]{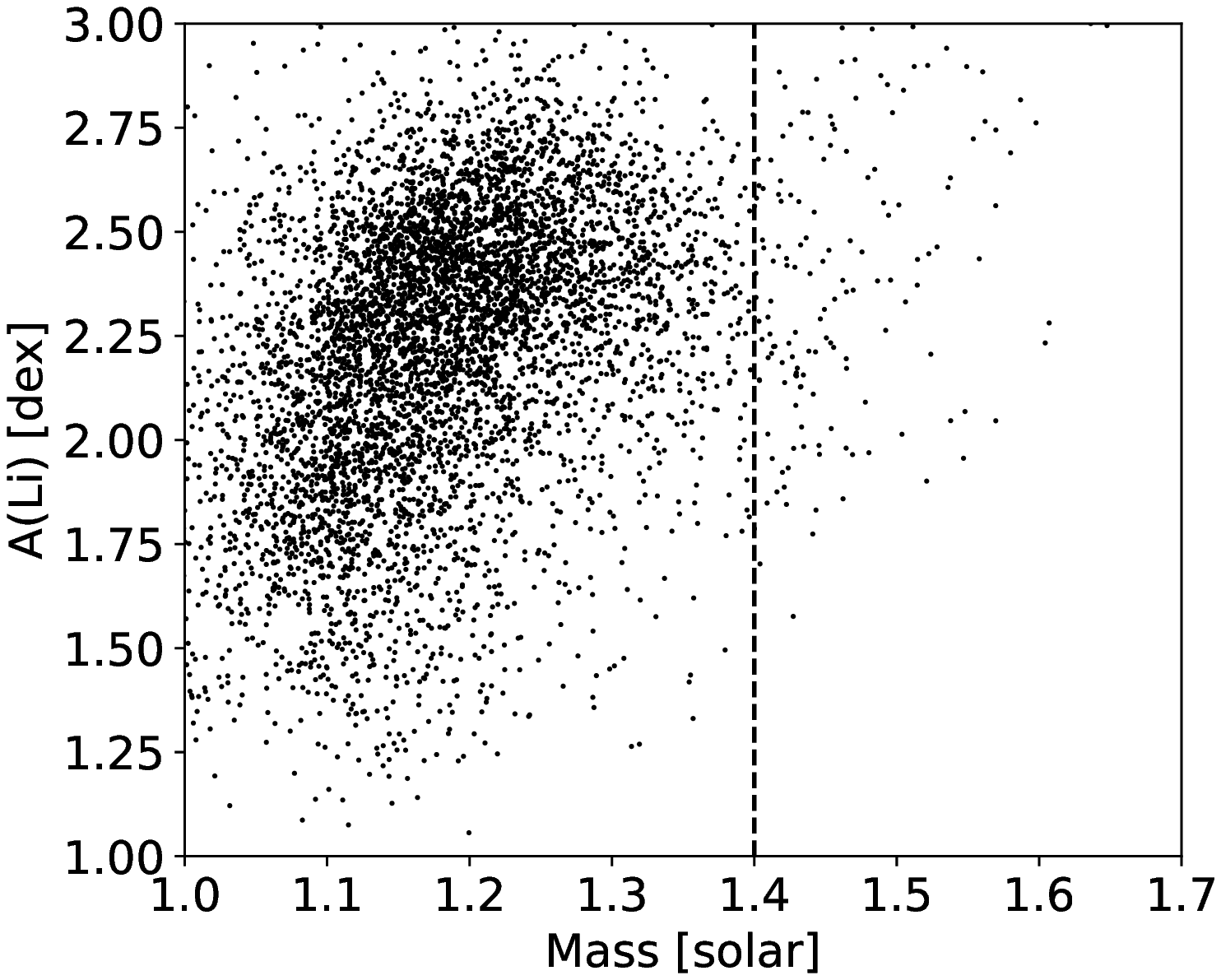}
    \caption{{Lithium abundance of stars in the GALAH survey as a function of stellar mass estimated from isochrone fitting. Stars shown have good quality flags, [Fe/H]$=0.2\pm0.2$ dex and $\log g \in [3.8,4.2]$ dex. The dashed line at mass of 1.4 \msun is shown for reference, this seems to represent the approximate location of the lithium dip at these metallicities. }} 
    \label{fig:galah_fig}
\end{figure}

\bibliographystyle{aasjournal} 
\bibliography{arxiv.bbl}
\end{document}